\documentclass[reprint,showkeys,showpacs,preprintnumbers,nofootinbib,amsmath,amssymb,aps,prd,
longbibliography,
onecolumn,
10pt]{revtex4}
\usepackage{array,multirow,graphicx}
\usepackage{dcolumn}
\usepackage{txfonts}
\usepackage{newlfont}
\usepackage{times}
\usepackage{bm}
\usepackage[colorlinks,citecolor=blue,urlcolor=blue,linkcolor=blue]{hyperref}
\usepackage[figtopcap]{subfigure}
\begin{document}
\title{Generic Phase Portrait Analysis of the Finite-time Singularities\\
and\\
Generalized Teleparallel Gravity}
\author{W. El Hanafy$^{1,3}$}%
\email{waleed.elhanafy@bue.edu.eg}
\author{G.G.L. Nashed$^{1,2,3}$}%
\email{nashed@bue.edu.eg}
\affiliation{$^{1}$Centre for Theoretical Physics, The British University in Egypt, P.O. Box 43, El Sherouk City, Cairo 11837, Egypt}
\affiliation{$^{2}$Mathematics Department, Faculty of Science, Ain Shams University, Cairo 11566, Egypt}
\affiliation{$^{3}$Egyptian Relativity Group (ERG), Cairo University, Giza 12613, Egypt}
\begin{abstract}
We analyze the common four types of the finite-time singularities using a generic framework of the phase portrait geometric approach. This technique requires that the Friedmann system to be written as a one dimensional autonomous system. We employ a scale factor that has been used widely in literature to realize the four finite-time singularity types, then we show a detailed discussion for each case showing possible novel models. Moreover, we show how different singularity types can play essential roles in different cosmological scenarios. Among several modified gravity theories, we show that the $f(T)$ cosmology is in comfort with the phase portrait analysis, since the field equations include Hubble derivatives only up to first order. Therefore, we reconstruct the $f(T)$ theory which generates these phase portraits. We also perform a complementary analysis using the effective equation of state. Furthermore, we investigate the role of the torsion fluid in realizing the cosmic singularities.
\end{abstract}
\maketitle
\section{Introduction}\label{Sec:1}
The cosmological singularity problem has received attentions longtime ago. However, lots of work were devoted for only the crashing type, e.g. big bang and big crunch singularities. Recently, after the supernovae of Type Ia (SNIa) observations, other softer types have been introduced to the game. Several attempts have been devoted to investigate the role of the quantum effects to smooth out the singularity \cite{Nojiri:2004ip,Tretyakov:2005en,Barrow:2011ub}, or to study the geodesic completeness and the possibilities of crossing the singularity \cite{Keresztes:2012zn,Barrow:2013ria,Kamenshchik:2016gcy,Awad:2017sau}. Also, some went to investigate the effect of the singularity on the cosmic observable quantities \cite{Dabrowski:2007ci,Keresztes:2009vc,Kleidis:2016vmd}. On the other hand, the nonsingular bounce cosmology could provide an alternative approach to explain cosmic observations. In this scenario, the universe begins with a contraction phase, then it reaches a nonzero minimal length before expansion, and therefore it does not suffer from trans-Planckian problems of the inflation models \cite{Martin:2000xs,Brandenberger:2012aj}. The bouncing universe has gained attentions in recent literature, whereas the cosmic evolution shows an interesting features on both the background and the perturbative levels \cite{Wands:1998yp,Finelli:2001sr,Biswas:2015kha,Novello:2008ra,Cai:2014bea,Battefeld:2014uga,Brandenberger:2016vhg}. In classical general relativity, one should introduce some fluid with an exotic equation of state to establish these models. However, we find that the modified gravity theories could provide an alternative by modifying the gravitational sector. Indeed, modified gravity has been used to describe bouncing universe successfully, which helps to resolve several problems of these models, e.g. anisotropy and ghost instability problems, c.f. \cite{Cai:2011tc,bounce3,Odintsov:2015ynk,Bamba2016,ElHanafy:2017sih}. Also, it is worth to analyzed various singularities within the modified gravity frame, which is the target of this paper.

For reasonable cosmological requirements, the universe is usually taken to be homogeneous and isotropic. Thus, we take the metric to be of the Friedmann-Lemaitre-Robertson-Walker (FLRW) form,
\begin{equation}\label{FLRW-metric}
ds^2=c^2 dt^{2}-a(t)^{2}\delta_{ij} dx^{i} dx^{j},
\end{equation}
where $c$ is the speed of light in vacuum and $a(t)$ is the scale factor of the universe. In modified theories of gravity, one can write the Friedmann system as
\begin{equation}\label{effective}
    \rho_{\textmd{eff}}\equiv \frac{3}{\kappa^2} H^{2},\quad p_{\textmd{eff}}\equiv -\frac{1}{\kappa^2}\left(2\dot{H}+3 H^{2}\right),
\end{equation}
where $H\equiv \frac{\dot{a}}{a}$ is the Hubble parameter, the dot denotes the derivative with respect to the cosmic time $t$, the coupling constant $\kappa^2\equiv 8 \pi G$, and $G$ is the gravitational constant. We use the natural units, i.e. $k_{B}=c=\hbar=1$. In this case, we can write $\kappa=1/M_{p}$, where $M_{p}=2.4 \times 10^{18}$ GeV is the reduced Planck mass. We also denote the effective (total) energy density and the pressure as $\rho_{\textmd{eff}}$ and $p_{\textmd{eff}}$, respectively. We further assume that the effective fluid having a barotropic equation of state, then we can classify the finite-time singularities as in Refs.~\cite{Nojiri:2005sx,Nojiri:2008fk}. In brief, we illustrate the fundamental characteristics of this classification as below: Let the time of the singularity be $t_s$,

(i) Type I: At $t \to t_s$, the scale factor $a$, the effective energy density $\rho_{\mathrm{eff}}$ and the pressure $p_\mathrm{eff}$ diverge, i.e. $a \to \infty$, $\rho_\mathrm{eff} \to \infty$, and $\left|p_\mathrm{eff}\right| \to \infty$. Type I singularity is of crushing type and known as ``\textit{big rip}" singularity, c.f. \cite{Nojiri:2005sx}.

(ii) Type II: At $t \to t_s$, the scale factor $a$ and the effective energy density $\rho_{\mathrm{eff}}$ approach finite values, while the effective pressure diverges, i.e. $a \to a_s$, $\rho_{\mathrm{eff}} \to \rho_s$ and $\left|p_\mathrm{eff}\right| \to \infty$. Type II singularity is not of a crushing type and known as ``\textit{sudden}" singularity, c.f. \cite{Nojiri:2004ip,Barrow:2004xh,Barrow:2004hk,Keresztes:2012zn,Kleidis:2016vmd}.

(iii) Type III : At $t \to t_s$, only the scale factor tends to a finite value, while the effective energy density and the pressure both diverge, i.e. $a \to a_s$, $\rho_\mathrm{eff} \to \infty$ and $\left|p_\mathrm{eff}\right| \to \infty$. Type III is of the crushing type.

(iv) Type IV : At $t \to t_s$, all the three mentioned quantities approach finite values, i.e. $a \to a_s$, $\rho_\mathrm{eff} \to \rho_s$ and $\left|p_\mathrm{eff}\right| \to p_s$. In addition, the Hubble parameter and its first derivative are finite, while its second/higher derivatives diverge. Type IV singularity is the softest (not of the crushing type) among the other four types, c.f. \cite{Nojiri:2005sx,Odintsov:2015ynk,Odintsov:2015gba,Odintsov:2015jca,Kleidis:2016vmd,Brevik:2016aos}.

The present work is devoted for analyzing the finite-time singularity types using a generic framework of the phase portrait geometric approach. This technique assumes that the Friedmann system to be written as a ``\textit{one dimensional autonomous system}" \cite{Awad2013,AHNS:2017}. So we organize the paper as follows: In Sec. \ref{Sec:2}, we review the phase portrait analysis and its application in cosmology. We also exhibit a scale factor that can realize the four singularity types, then we show a detailed discussion for each case. Moreover, we show that how different singularity types can play essential roles in different cosmological scenarios. In Sec. \ref{Sec:3}, we show that the $f(T)$ gravity among several modified gravity theories is in comfort with the phase portrait technique. We also reconstruct the $f(T)$ theory which generates these phase portraits. In Sec. \ref{Sec:4}, we perform a complementary analysis using the effective equation of state. Furthermore, we investigate the role of the torsion fluid in realizing the cosmic singularities. In Sec. \ref{Sec:5}, we summarize the work.
\section{Phase portraits of finite-time singularities}\label{Sec:2}
In Eqs. (\ref{effective}), we define the effective density energy and the effective pressure as $\rho_{\textmd{eff}}=\mathop{\sum}\limits_{i}\rho_{i}$ and $p_{\textmd{eff}}=\mathop{\sum}\limits_{i}p_{i}$, and the index $i$ denotes the fluid component. In this case, if we consider the effective matter to have a linear equation of state, we define the effective equation of state parameter as
\begin{equation}\label{eff_EoS0}
\omega_{\textmd{eff}} \equiv \frac{p_{\textmd{eff}}}{\rho_{\textmd{eff}}}=-1-\frac{2}{3}\frac{\dot{H}}{H^2}.
\end{equation}
In fact, the differential equation represents a one dimensional autonomous system, if we can write $\dot{H}\equiv \mathcal{F}(H)$. The proper way to fulfill this condition is to have a fixed effective equation of state parameter which reproduces the general relativistic as a special case, or to have a dynamical effective equation of state but as a function of $H$ only. We are interested in the more general case, $\omega_{\textmd{eff}}\equiv \omega_{\textmd{eff}}(H)$, so rewrite (\ref{eff_EoS0}) as
\begin{equation}\label{eff_EoS1}
\dot{H}=-\frac{3}{2}\left(1+\omega_{\textmd{eff}}(H)\right)H^{2}\equiv \mathcal{F}(H).
\end{equation}
As a consequence, we can always interpret this differential equation as a vector field on a line introducing one of the basic techniques of dynamics. This view, geometrically, is by drawing $\dot{H}$ verses $H$, which helps to analyze the cosmic model in a clear and transparent way even without solving the system. In order to fix our notations, we follow \cite{book:Steven} calling equation (\ref{eff_EoS1}) the \textit{phase portrait}, while its solution $H(t)$ is the \textit{phase trajectory}. Thus, the phase portrait corresponds to any theory can be drawn in an ($\dot{H}-{H}$) \textit{phase-space} of the Friedmann's system. In this space each point is a \textit{phase point} and could serve as an initial condition.

\begin{figure}
\centering
\includegraphics[scale=.35]{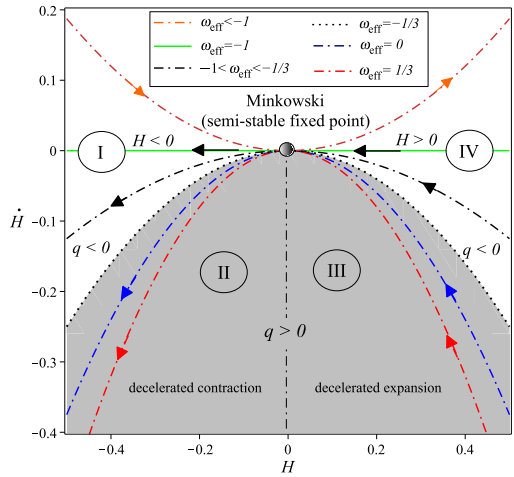}
\caption{Phase portraits (\ref{eff_EoS1}) corresponding to different values of the effective equation of state parameters. The origin of the phase space is the Minkowski space, and the zero acceleration curve ($\omega_{\textmd{eff}}=-1/3$) splits the space into deceleration (shaded) and acceleration (unshaded) regions, while the negative (positive) Hubble regions are contraction (expansion) regions. This subdivides the phase space into four regions, we denote them accordingly as I$-$IV regions. }
\label{Fig:GR-phase-portraits}
\end{figure}

For $C^{\infty}$ monotonic phase portraits, it has been shown that the fixed points ($\dot{H}=0$) cannot be reached in a finite time \cite{Awad2013}. In fact, the fixed point, in addition, would be Minkowskian, if $H=0$ and $\dot{H}=0$. Otherwise, it would be a de Sitter, i.e. $H\neq 0$ and $\dot{H}=0$. For a phase portrait (\ref{eff_EoS1}), if the universe begins at a fixed point, it will stay forever at that point. The stability classification of a fixed point can be performed by allowing a small perturbation about it, and investigating that whether the perturbation decays or grows. Generally, if the slope of the phase portrait is positive at the fixed point, the perturbation grows and the fixed point in this case is unstable (repeller or source). If the slope is negative, the fixed point is stable (attractor or sink). However, if the slop alters its sign at the fixed point, it is semi-stable and the solution is stable from one side and unstable from the other side. For an infinite slope case (e.g. cusps) there are infinitely many solutions starting from the same initial condition so that the phase point doesn't know how to move. In this case the geometric approach collapses.\\

We summarize the phase portrait analysis as a useful tool to qualitatively describe the dynamical behavior of a flat FLRW model by constructing its ($\dot{H}\textmd{-}H$) phase space diagram \cite{Awad2013}. When the system is an autonomous one dimensional, the dynamics would be controlled by the asymptotic behavior of the phase portrait and by its fixed points. In general, after identifying the fixed points, we next define the zero acceleration curve by setting the deceleration parameter $q\equiv -\frac{a\ddot{a}}{\dot{a}^{2}}=0$, i.e., $\dot{H}=-H^2$, which splits the phase space into two regions. The inner region characterizes the decelerated phases, this is shown by the shaded region in Fig. \ref{Fig:GR-phase-portraits}. However, the unshaded region represents the accelerated phases. We classify different phases in Fig. \ref{Fig:GR-phase-portraits} as follows:

(i) Region I: it represents an accelerated contracting universe as $q<0$ and $H<0$.

(ii) Region II: it represents a decelerated contacting universe as $q>0$ and $H<0$.

(iii) Region III: it represents a decelerated expanding universe as $q>0$ and $H>0$ which characterizes the usual FLRW models.

(iv) Region IV: it represents an accelerated expanding universe as $q<0$ and $H>0$ which characterizes the so-called inflation or dark energy phases.

We consider the following scale factor in order to realize the four singularity types \cite{Odintsov:2015zza,Odintsov:2015gba}
\begin{equation}\label{scale.factor}
    a(t)=e^{f_{0}(t-t_{s})^{2(1+\varepsilon)}},
\end{equation}
Introducing the new parameters $\alpha\equiv 1+2\varepsilon$ and $\beta\equiv 2(1+\varepsilon)f_{0}=(1+\alpha)f_{0}$, we write the Hubble parameter as
\begin{equation}\label{Hubble}
    H(t)=\beta(t-t_{s})^{\alpha},
\end{equation}
where $\alpha$ is dimensionless and $\beta$ has a dimension of [time]$^{-1-\alpha}$, we note that in the Planck unit system, the Hubble parameter is measured in [GeV] and the time is measured in [GeV]$^{-1}$, then $\beta$ is measured in [GeV]$^{1+\alpha}$. The classification of finite-time singularities can be realized by the scale factor (\ref{scale.factor}). According to different choices of the parameter $\alpha$, as given in Table \ref{Table:classification}, the four types of singularities could occur at $t=t_s$.
\begin{table}
\caption{Singularity types classification according to the choice of the $\alpha$ parameter in Eq. (\ref{scale.factor})}
\begin{center}\begin{tabular}{cccccc}
\hline\hline
\multirow{2}{*}{$t \to t_{s}$}& Type I& Type II& Type III & Type IV$^{*}$\\
             & ($\alpha<-1$) & ($0<\alpha<1$)& ($-1<\alpha<0$)& ($\alpha>1$)\\
\hline
$a \to a_{s}$& $\times$& $\checkmark$& $\checkmark$ & $\checkmark$\\
$H \to H_{s}$& $\times$& $\checkmark$& $\times$ & $\checkmark$\\
$\dot{H} \to \dot{H}_{s}$& $\times$& $\times$ & $\times$ & $\checkmark$\\
\hline
$\rho_{\textmd{eff}} \to \rho_{s}$& $\times$& $\checkmark$ & $\times$ & $\checkmark$\\
$|p_{\textmd{eff}}| \to p_{s}$& $\times$ & $\times$& $\times$ & $\checkmark$\\
  \hline
\hline
\end{tabular}\end{center}
\noindent{$^{*}$The higher derivatives $d^{n}H/dt^{n}$ diverge, $n\geq 2$.}
\label{Table:classification}
\end{table}
Using the inverse relation of Eq. (\ref{Hubble}), we express the time in terms of Hubble,
\begin{equation}\label{t(H)}
    t(H)=t_{s}+\left(\frac{H}{\beta}\right)^{\frac{1}{\alpha}}.
\end{equation}
Thus, we write the implicit differentiation $\dot{H}\equiv \dot{H}(H)$ as
\begin{equation}\label{autonomous}
    \dot{H}=\alpha H \left(\frac{H}{\beta}\right)^{-\frac{1}{\alpha}}.
\end{equation}
The above relation represents a one dimensional autonomous system, whereas its graphical representation provides the \textit{phase portrait} of the cosmic evolution, which can be seen clearly in the $\left(H \textmd{-} \dot{H}\right)$ phase space for different choices of the parameter $\alpha$.

In general, the phase portrait (\ref{autonomous}) evolves towards a fixed point at which $\dot{H}=0$, or a point at which $\dot{H}\to \pm \infty$. The former can be achieved at $H=0$ where $\alpha<0$ or $\alpha>1$ so we expect that the phase portraits associated with Type I, III and IV singularities to pass through a Minkowskian fixed point $\left(H=0,~\dot{H}=0\right)$. The later does not always mean that the system is singular at $\dot{H}\to \pm \infty$. It has been shown that if the asymptotic behavior of $\dot{H}$ grows linearly or slower, the solution will not have a finite-time singularity \cite{Awad2013}. This can be shown, since the singularity time
\begin{equation}\label{sing-time}
t_s=\int_{H_0}^{\pm \infty} \frac{dH}{\dot{H}}.
\end{equation}
Using Eq. (\ref{autonomous}), it is clear that the phase portrait belongs to the power-law family, i.e. $\dot{H}\propto H^{\gamma}$. It is not difficult to realize that the big bang (singular) model is a special case where $\gamma=2$. In general, the time to reach the singularity is finite
$$t_s=\frac{1}{(\gamma-1)H_{0}^{\gamma-1}};\quad \textmd{if }\gamma>1.$$
On the contrary, the time to reach the singularity could be infinite
$$t_s\to \pm \infty; \quad \textmd{if } \gamma\leq 1.$$
Therefore, the phase portrait (\ref{autonomous}) evolves towards $\dot{H}\to \pm \infty$ in an infinite time where $\alpha>0$, and we expect that the phase portraits associated with Type II and IV singularities only to be free from singularities of crashing type at $H\to \pm \infty$.
\subsection{Type I singularity phase portrait}\label{Sec:2.1}
This singularity occurs when the cosmic time approaches $t \to t_s$, the scale factor $a$, the effective energy density $\rho_{\mathrm{eff}}$ and the pressure $p_\mathrm{eff}$ diverge, i.e. $a \to \infty$, $\rho_\mathrm{eff} \to \infty$, and $\left|p_\mathrm{eff}\right| \to \infty$. Type I singularity is of crushing type and known as ``\textit{big rip}" singularity, c.f. \cite{Nojiri:2005sx}. Using the scale factor (\ref{scale.factor}), the Type I singularity case occurs when $\alpha < -1$. In this case, we have $\beta=f_0(1+\alpha)<0$, when $f_{0}>0$ and $\beta=f_0(1+\alpha)>0$, when $f_{0}<0$. Different cases of the phase portraits corresponding to Eq. (\ref{autonomous}) are given in Fig. \ref{Fig:phspTypeI}. In general, the one dimensional autonomous systems are dominated by the fixed points, whereas the time required to reach any of these points is infinite. On another word, the fixed points split the phase space to separate patches. This fact can be seen easily in Friedmann cosmology, since the time is given as
\begin{equation}\label{time:fixed-point}
t=\int_{H_{0}}^{H_{f}}\frac{dH}{\dot{H}} \to \pm \infty.
\end{equation}

At a fixed point $H_{f}$, we have $\dot{H}=0$; and the time required to reach it is infinite. The phase portraits in Fig. \ref{Fig:phspTypeI} show two possible cosmic behaviors according to the choice of $f_{0}>0$ ($f_{0}<0$), consequently $\beta<0$ ($\beta>0$). So we discuss both cases as follows:
\subsubsection{$f_{0}>0$ ($\beta<0$)}\label{Sec:2.1.1}
\begin{figure}
\centering
\subfigure[~$f_{0}>0$ ($\beta<0$)]{\label{fig:phsptypeIa}\includegraphics[scale=.3]{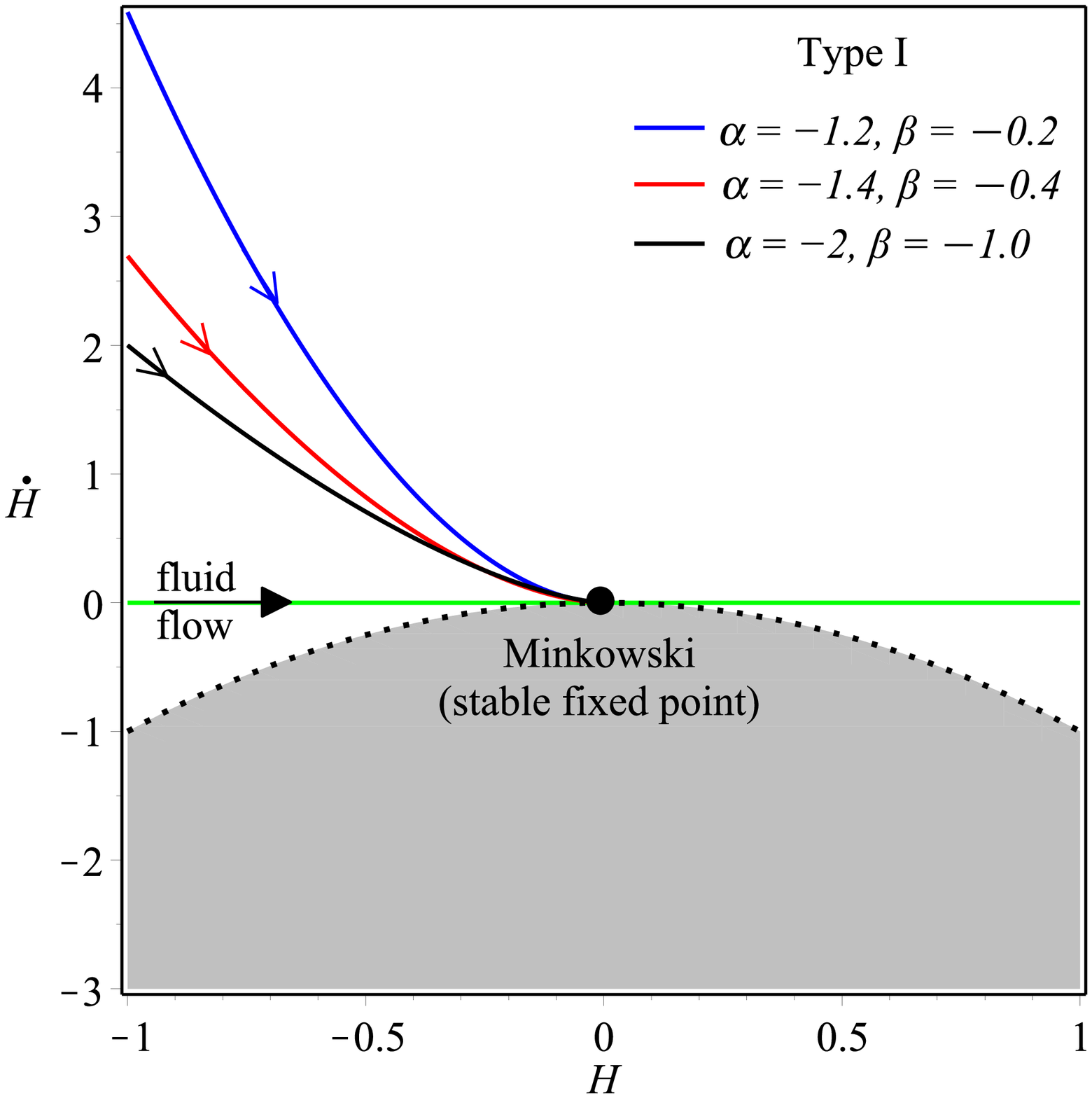}}
\subfigure[~$f_{0}<0$ ($\beta>0$)]{\label{fig:phsptypeIb}\includegraphics[scale=.34]{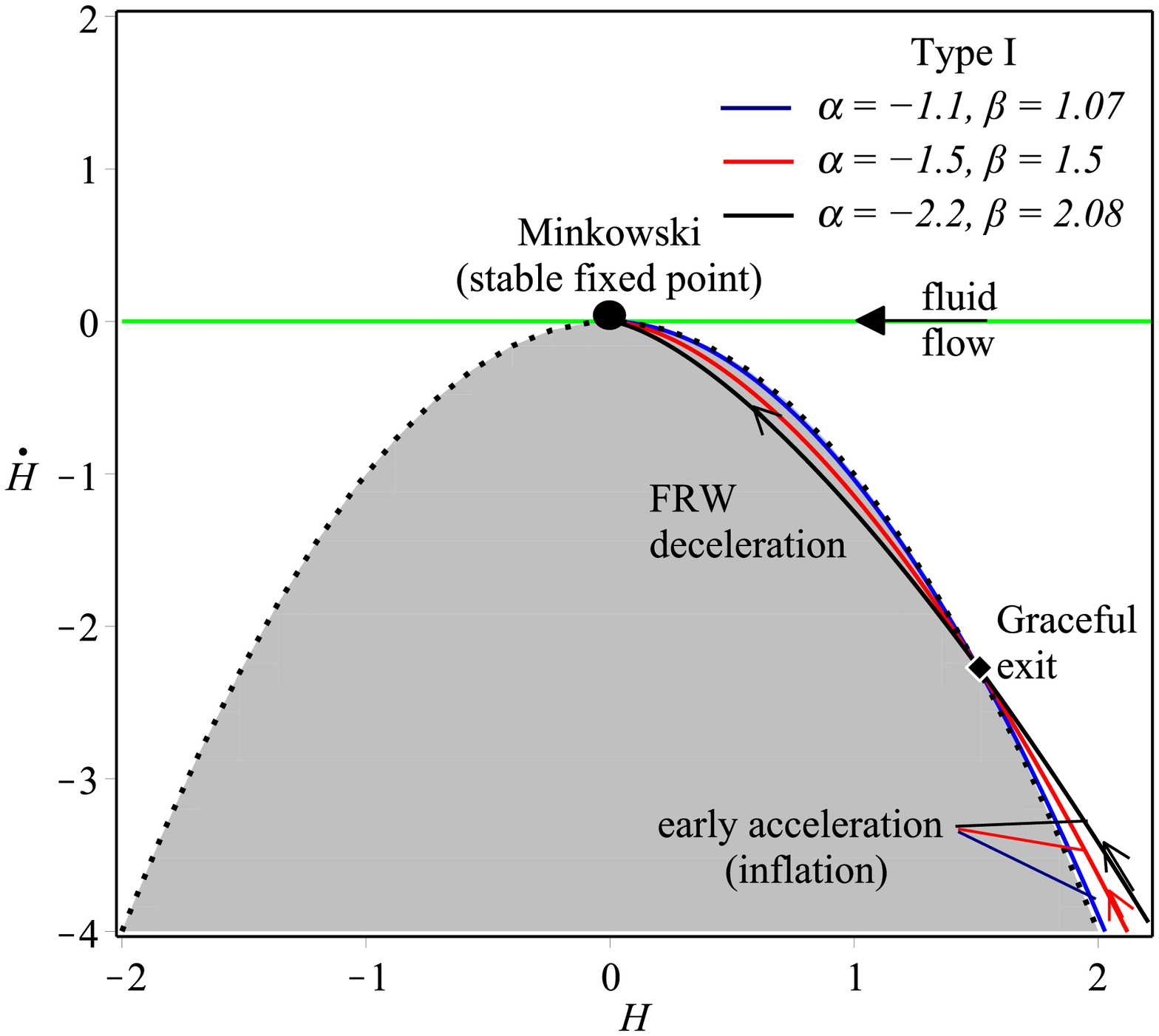}}
\caption[figtopcap]{The phase portraits of the finite-time singularities of Type I for different choices of the parameter $\alpha<-1$ in Eqs. (\ref{autonomous}).}
\label{Fig:phspTypeI}
\end{figure}

The corresponding phase portrait is given on Fig. \ref{Fig:phspTypeI}\subref{fig:phsptypeIa}. The universe is allowed only in the phantom regime, it begins with a big rip singularity at
$$t_{s}=\int_{-\infty}^{H_{0}}\frac{dH}{\dot{H}}=\left|\frac{H_{0}}{\beta}\right|^{1/\alpha},$$
from a particular value $H_{0}<0$. Since the universe evolves in $\dot{H}>0$ and $H<0$ phase space with a Minkowskian fixed point fate, the universe experiences an eternal accelerating contraction.
\subsubsection{$f_{0}<0$ ($\beta>0$)}\label{Sec:2.1.2}
The corresponding phase portrait is given on Fig. \ref{Fig:phspTypeI}\subref{fig:phsptypeIb}. The universe evolves in a non-phantom regime, it begins with a big bang at $t_{s}=(H_{0}/\beta)^{1/\alpha}$ from a present Hubble value $H_{0}>0$. Interestingly, for suitable values of the parameters $\alpha$ and $\beta$, the phase portrait provides a graceful exit inflationary model. The universe could begin with an early accelerated expansion phase (inflation), however, it could be followed by a decelerated expansion phase (FLRW). At the graceful exit transition, the phase portrait should cut the zero acceleration curve ($\dot{H}=-H^{2}$) from acceleration to deceleration. Using (\ref{autonomous}), at the transition $H_{inf}$, the two parameters can be related by
\begin{equation}\label{inf-end}
    \beta=H_{inf}\left(-\frac{H_{inf}}{\alpha}\right)^{\alpha}.
\end{equation}
For any value of $\alpha<-1$, the above equation can predict the value of $\beta$ when the Hubble parameter at the end of inflation $H_{inf}$ is known. For more detailed discussion about the possible values of $\alpha$ and $\beta$ to realize a graceful exit inflation, see Sec. \ref{Sec:2.5}.
\subsubsection{the physical description}
In conclusion, we find that the case of $\beta>0$ is more interesting to be studied among the other cases, since it provides a graceful exit inflation model. In Sec. \ref{Sec:2.5}, we will obtain the suitable values of the parameters $\alpha$ and $\beta$ which realize inflation at a suitable energy scale.
\subsection{Type II singularity phase portrait}\label{Sec:2.2}
This singularity occurs when the cosmic time approaches $t \to t_s$, the scale factor $a$ and the effective energy density $\rho_{\mathrm{eff}}$ approach finite values, while the effective pressure diverges, i.e. $a \to a_s$, $\rho_{\mathrm{eff}} \to \rho_s$ and $\left|p_\mathrm{eff}\right| \to \infty$. Type II singularity is not of a crushing type and known as ``\textit{sudden}" singularity, c.f. \cite{Nojiri:2004ip,Barrow:2004xh,Barrow:2004hk,Keresztes:2012zn,Kleidis:2016vmd}. Using the scale factor (\ref{scale.factor}), the Type II singularity case occurs when $0 <\alpha < 1$. In this case, we have $\beta=f_0(1+\alpha)<0$, when $f_{0}<0$, while $\beta=f_0(1+\alpha)>0$, when $f_{0}>0$. Different cases of the phase portraits corresponding to Eq. (\ref{autonomous}) are given in Fig. \ref{Fig:phspTypeII}, it is clear that the finite-time singularity of type II is not a fixed point. But it occurs, commonly, when $\dot{H}$ diverges as $H \to H_s=0$. However, we split the discussion of the phase portraits corresponding to the Type II singularity into two main categories according to the choice of $f_{0}<0$ and $f_{0}>0$ in the given range $0<\alpha<1$ as follows:
\begin{figure}
\centering
\subfigure[~$\alpha=\frac{1}{n}$, $n$ = odd]{\label{fig:phsptypeIIa}\includegraphics[scale=.3]{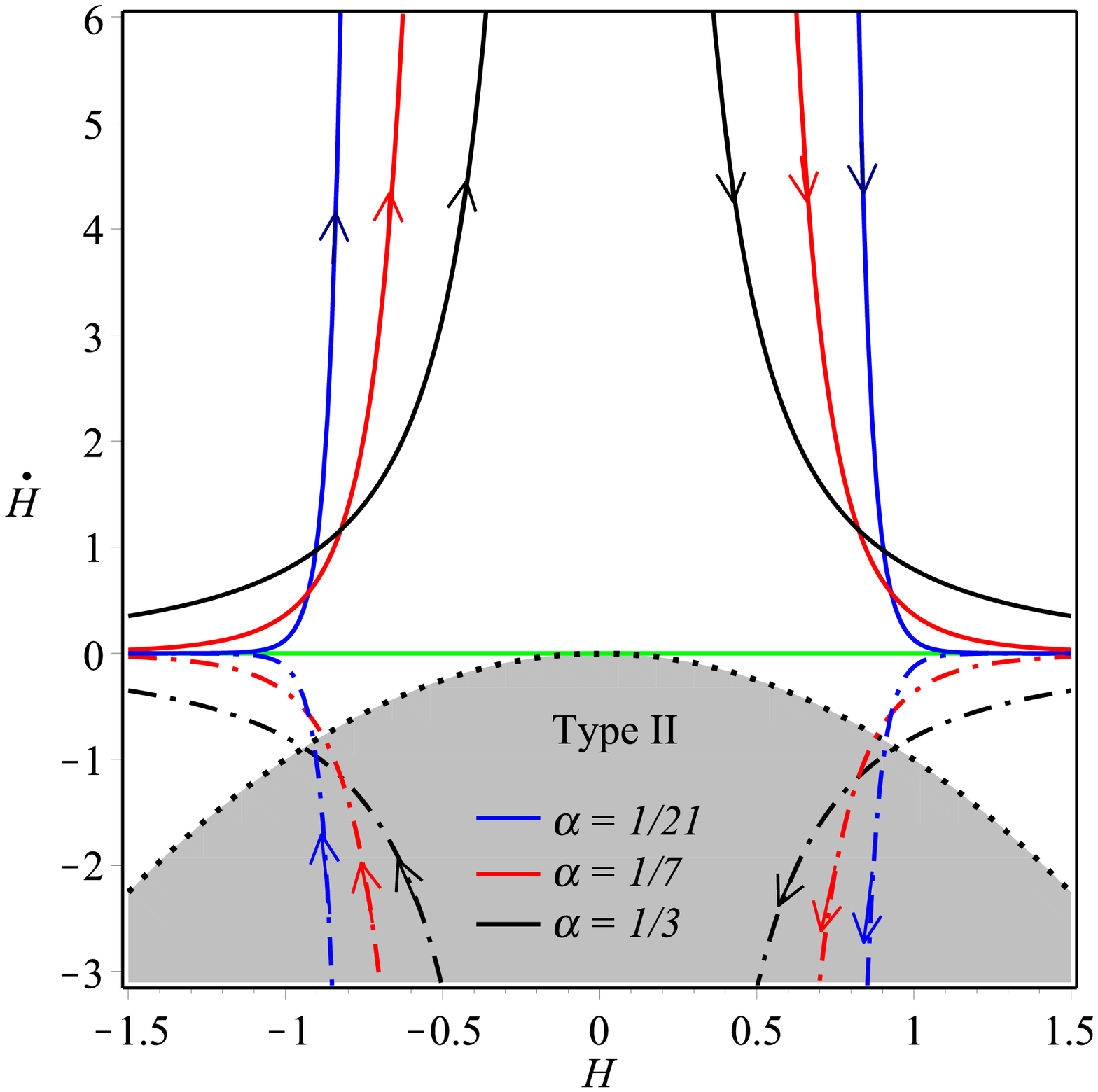}}
\subfigure[~$\alpha=\frac{1}{n}$, $n$ = even]{\label{fig:phsptypeIIb}\includegraphics[scale=.3]{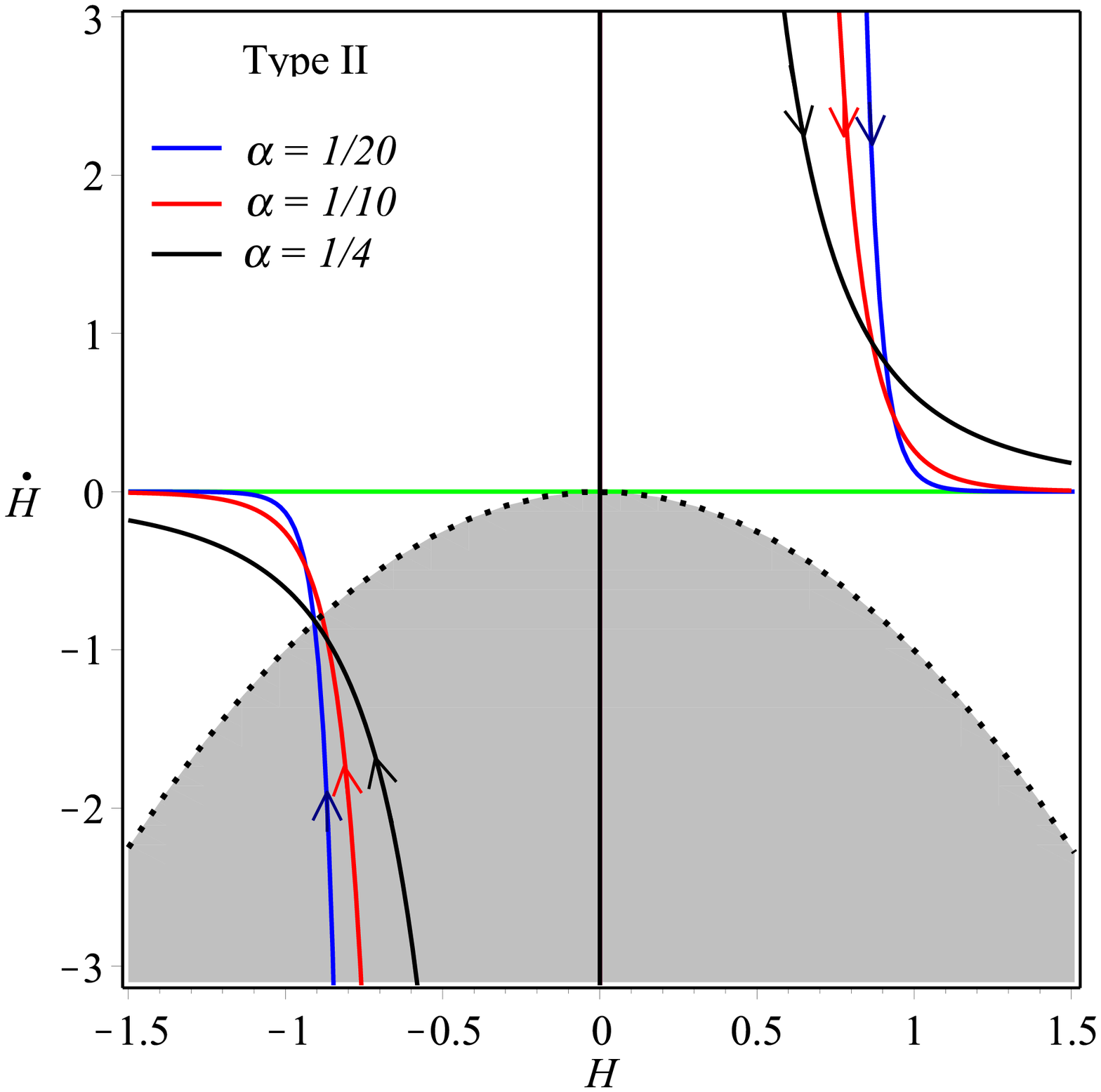}}
\subfigure[~$\alpha=\frac{a}{b}$, $a<b$ and $\frac{a}{b}\neq\frac{1}{n}$]{\label{fig:phsptypeIIc}\includegraphics[scale=.3]{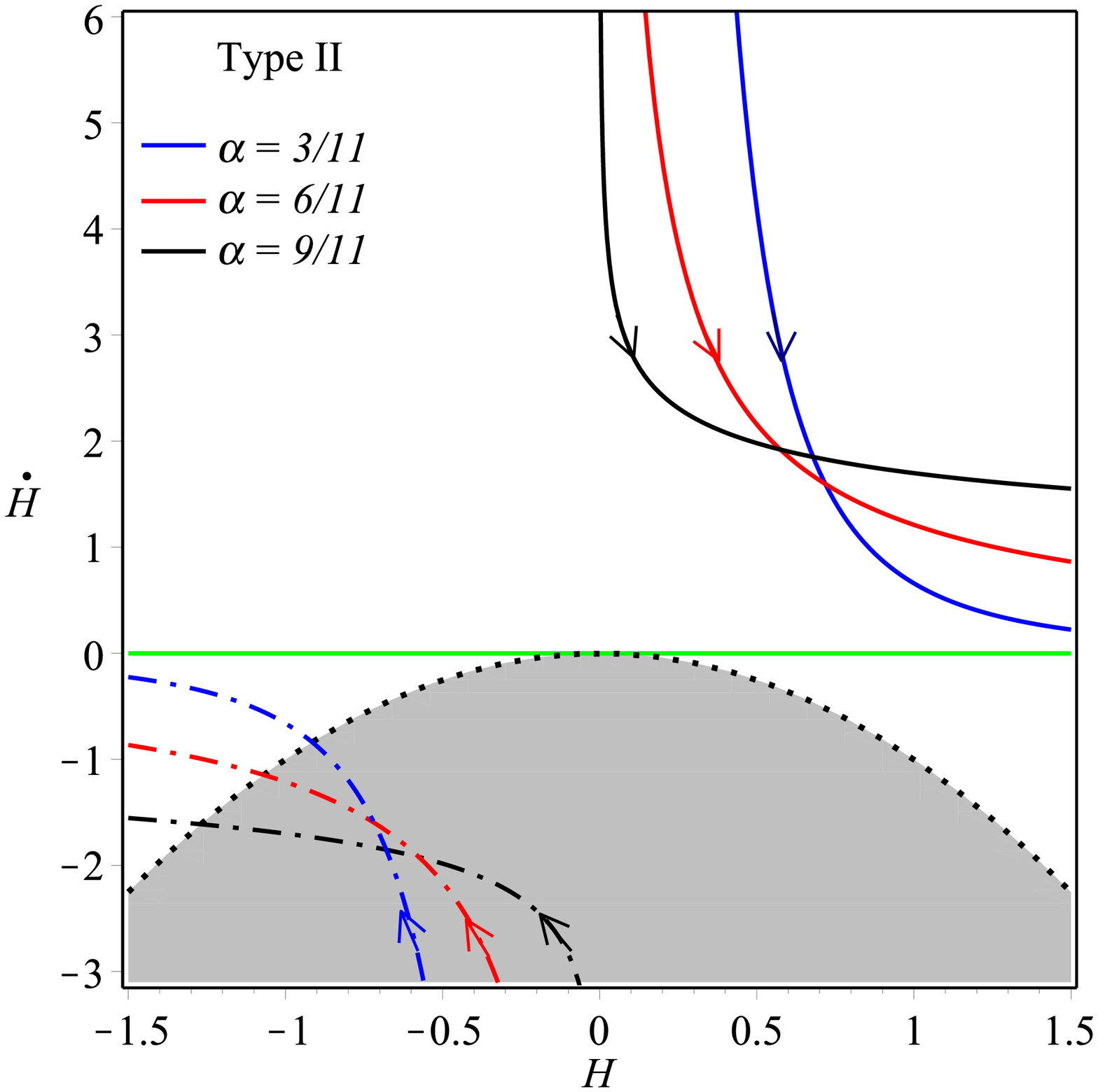}}
\caption[figtopcap]{The phase portraits of the finite-time singularities of Type II for different choices of the parameter $0<\alpha<1$ in Eqs. (\ref{autonomous}). The dash curves are corresponding to $f_0<0$ ($\beta<0$), and the solid curves are corresponding to $f_0>0$ ($\beta>0$). Same color curves means we have used the same value of $\alpha$. The phase portraits of Fig. \subref{fig:EoS-typeIIb} are typical for both cases $\beta<0$ and $\beta>0$.}
\label{Fig:phspTypeII}
\end{figure}
\subsubsection{$f_{0}<0$ ($\beta<0$)}\label{Sec:2.2.1}
In this category, three patterns could be obtained according to the value of $\alpha$. Therefore, we have the following subclasses:\\

\textit{Case 1.} $\alpha=\frac{1}{n}$ ($n>1$ is an odd positive integer): The corresponding phase portraits are indicated by the dash curves as in Fig. \ref{Fig:phspTypeII}\subref{fig:phsptypeIIa}. At the positive $H$ region, the fluid evolves towards left (decreasing $H$) approaching a future sudden singularity $H_s=0$, where the time to reach it is finite.
$$t_s=\int_{H_0}^{H_s=0}\frac{dH}{\dot{H}}=-\left(\frac{H_{0}}{\beta}\right)^{1/\alpha},$$
where $H_{0}>0$ denotes the present Hubble value. At the negative $H$ region, the cosmic fluid evolves towards left (decreasing $H$). It begins with a sudden singularity $\dot{H}\to -\infty$ at $H_s=0$, then it goes in a decelerated contraction phase. After that it enters an accelerated contraction phase with a de Sitter fate as $H\to -\infty$. Similar models have been studied in details \cite{Keresztes:2012zn}.\\

\textit{Case 2.} $\alpha=\frac{1}{n}$ ($n>1$ is an even positive integer): The corresponding phase portraits are given on Fig. \ref{Fig:phspTypeII}\subref{fig:phsptypeIIb}. Notably, in this subclass, the phase portraits corresponding to $\pm~ \beta$ are identical. At the negative $H$ region, the cosmic fluid flows towards the left direction (decreasing $H$). However, at the positive $H$ region, the fluid flows towards the right direction (increasing $H$). So we may consider these two regions as separate patches, the left patch ($H<0$) begins with a finite-time singularity of type II, whereas the universe evolves from a decelerated contraction to an accelerated expansion towards a de Sitter space. On the contrary, the right patch ($H>0$) shows that the universe has an initial finite-time singularity of type II, but evolves effectively towards a de Sitter space in a phantom regime.\\

\textit{Case 3.} $\alpha=\frac{a}{b}$ ($a<b$ and $\frac{a}{b}\neq \frac{1}{n}$): The corresponding phase portraits are indicated by the dash curves as in Fig. \ref{Fig:phspTypeII}\subref{fig:phsptypeIIc}, which shows that the positive $H$ patch is not valid for these values of $\alpha$. However, the dynamical behavior of the universe is the same as in cases 1 and 2 above, in the $H<0$ patch.
\subsubsection{$f_{0}>0$ ($\beta>0$)}\label{Sec:2.2.2}
In this category, similarly, three patterns could be obtained according to the value of $\alpha$. Therefore, we have the following subclasses:\\

\textit{Case 1.} $\alpha=\frac{1}{n}$ ($n>1$ is an odd positive integer): The corresponding phase portraits are indicated by the solid curves as in Fig. \ref{Fig:phspTypeII}\subref{fig:phsptypeIIa}. At the negative $H$ region (contraction phase), the cosmic fluid flows towards the right direction (increasing $H$). For a given value $H_{0}<0$, the universe has no initial singularity as
$$t=\int_{-\infty}^{H_{0}}\frac{dH}{\dot{H}}=-\infty.$$
However, the time required to reach the singular phase is
$$t_{s}=\int_{H_{0}}^{0}\frac{dH}{\dot{H}}=-\left|\frac{H_{0}}{\beta}\right|^{1/\alpha}.$$
At the positive $H$ region (expansion phase), the fluid flows , also, towards the right direction. The time from the singular phase $H=0$ to a given phase $H_{0}>0$ is $t_{s}=\left(\frac{H_{0}}{\beta}\right)^{1/\alpha}$, while the universe has no future singularity as the time to reach $H \to \infty$ is infinite. Since $\dot{H}>0$ at both regions, $H<0$ and $H>0$, and time is extended as $-\infty< t< \infty$, this case gives rise to a bouncing model with a singularity of type II at the bouncing point $H=0$. In this case, we can restrict the Hubble length to a minimal value of the Planck length and use junction conditions to cross the singularity, providing a soft rebirth of the expanding universe after the contraction phase \cite{Keresztes:2012zn}.\\

\textit{Case 2.} $\alpha=\frac{1}{n}$ ($n>1$ is an even positive integer): The corresponding phase portraits are given on Fig. \ref{Fig:phspTypeII}\subref{fig:phsptypeIIb}. As mentioned in Sec. \ref{Sec:2.2.1}, the phase portraits of $\pm~ \beta$ are identical. Therefore, we expect the same description as given before.\\

\textit{Case 3.} $\alpha=\frac{a}{b}$ ($a<b$ and $\frac{a}{b}\neq \frac{1}{n}$): The corresponding phase portraits are indicated by the solid curves as in Fig. \ref{Fig:phspTypeII}\subref{fig:phsptypeIIc}, which shows that the negative $H$ patch is not valid for these values of $\alpha$. However, the dynamical behavior of the universe is the same as in cases 1 and 2 above, in the $H>0$ patch.\\
\subsubsection{the physical description}\label{Sec:2.2.3}
In conclusion, the finite-time singularity of type II could act as an attractor or repeller according to the choice of the model parameter in the given range $0<\alpha<1$. Remarkably, case 1, for $\beta>0$ values, provides a big bounce cosmology in the phantom regime, where the bouncing point coincides with a sudden singularity. Also, case 1, for $\beta<0$ values, provides a big brake cosmology, this has been shown to be compatible with the SNIa data \cite{Keresztes:2012zn}. In both cases, the first derivative of the scale factor is finite, so the Christoffel symbols are regular and the geodesics are well behaved, and thus the singularity becomes traversable. In addition to the kinematical description of phase portrait method, a complementary investigation using the torsion gravity is available in Sec. \ref{Sec:4.2}.
\subsection{Type III singularity phase portrait}\label{Sec:2.3}
This singularity occurs when the cosmic time approaches $t \to t_s$, only the scale factor tends to a finite value, while the effective energy density and the pressure both diverge, i.e. $a \to a_s$, $\rho_\mathrm{eff} \to \infty$ and $\left|p_\mathrm{eff}\right| \to \infty$. Type III is of the crushing type. Using the scale factor (\ref{scale.factor}), the Type III singularity case occurs when $-1 <\alpha < 0$. In this case, we have $\beta=f_0(1+\alpha)<0$, when $f_{0}<0$, while $\beta=f_0(1+\alpha)>0$, when $f_{0}>0$. Different cases of the phase portraits corresponding to Eq. (\ref{autonomous}) are given in Fig. \ref{Fig:phspTypeIII}. Remarkably, the transition from deceleration to acceleration, in the $H>0$ region, can be realized for $-1<\alpha<0$. Since the evolution pattern is sensitive to the choices of $\alpha$ and $\beta$, we discuss different cases as follows:
\begin{figure}
\centering
\subfigure[~$\alpha=-\frac{1}{n}$, $n$ = odd]{\label{fig:phsptypeIIIa}\includegraphics[scale=.4]{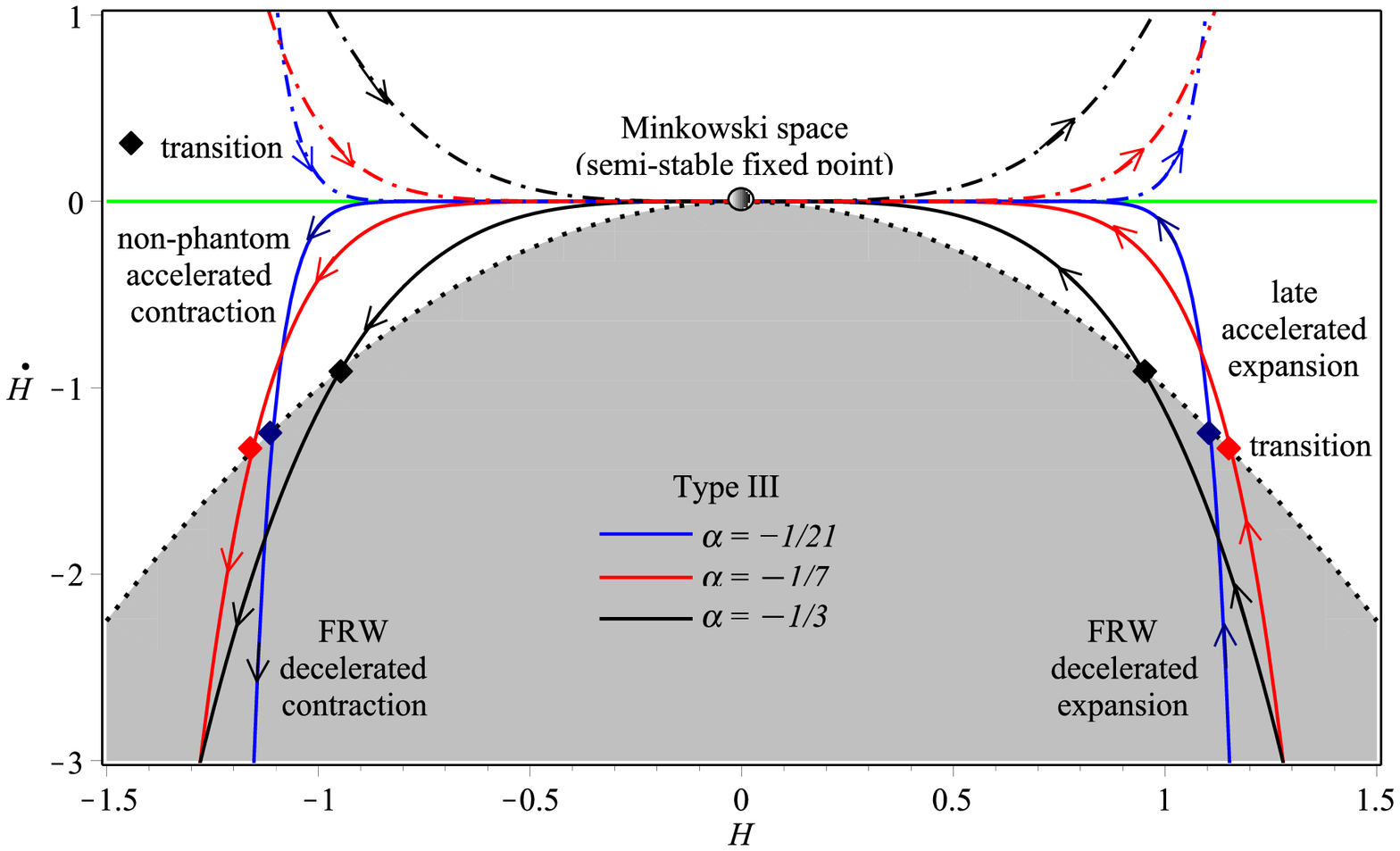}}
\subfigure[~$\alpha=-\frac{1}{n}$, $n$ = even]{\label{fig:phsptypeIIIb}\includegraphics[scale=.4]{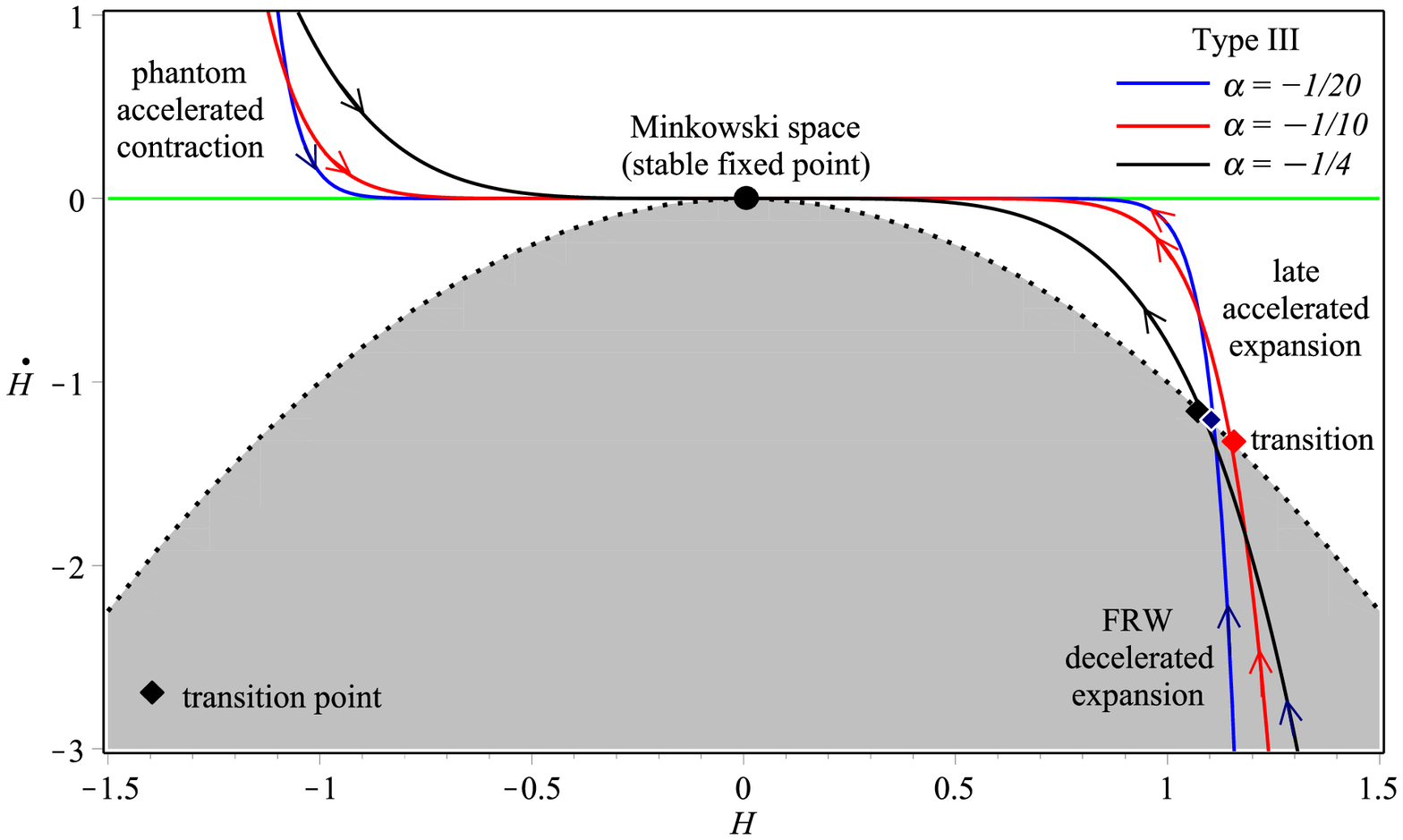}}\\
\subfigure[~$\alpha=-\frac{a}{b}$, $a<b$ and $\frac{a}{b}\neq\frac{1}{n}$]{\label{fig:phsptypeIIIc}\includegraphics[scale=.4]{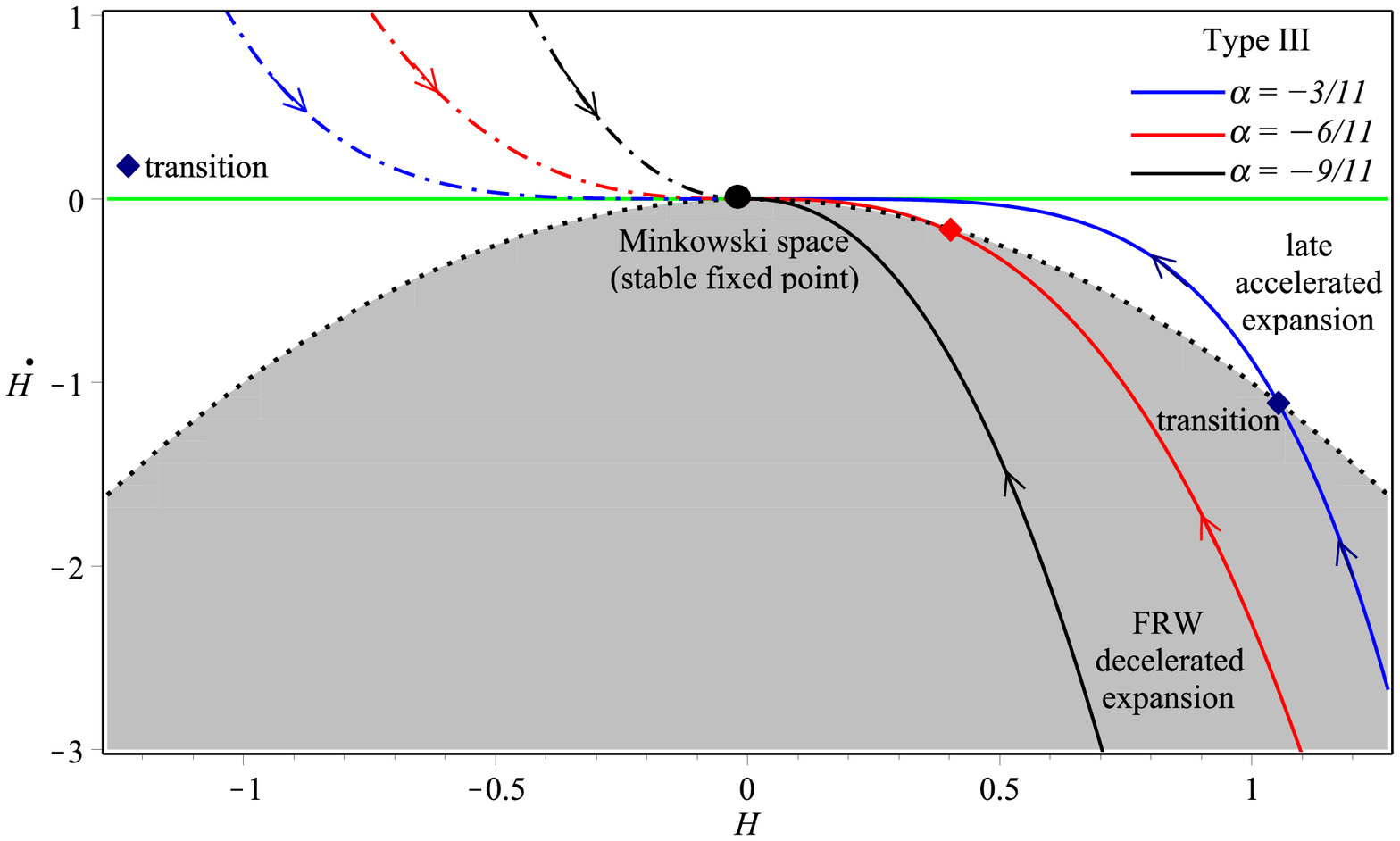}}
\caption[figtopcap]{The phase portraits of the finite-time singularities of Type III for different choices of the parameter $-1<\alpha<0$ in Eqs. (\ref{autonomous}). The dash curves are corresponding to $f_0<0$ ($\beta<0$), the solid curves are corresponding to $f_0>0$ ($\beta>0$). Same color curves means we have used the same value of $\alpha$. The phase portraits of Fig. \subref{fig:EoS-typeIIIb} are typical for both cases $\beta<0$ and $\beta>0$.}
\label{Fig:phspTypeIII}
\end{figure}
\subsubsection{$f_0<0$ ($\beta<0$)}\label{Sec:2.3.1}
In this category, three patterns could be obtained according to the value of $\alpha$. Therefore, we have the following subclasses:\\

\textit{Case 1.} $\alpha=-\frac{1}{n}$ ($n>1$ is an odd positive integer): The phase portraits for some particular choices of $\alpha$ are indicated by the dash curves as in Fig. \ref{Fig:phspTypeIII}\subref{fig:phsptypeIIIa}, where the Mikowskian fixed point at the origin of the phase space is a common point for all portraits. The Minkowski space, in this case, is a semi-stable fixed point. Since the universe requires an infinite time to reach this Minkowskian space, we find that the regions, $H<0$ and $H>0$, are two separate patches. Firstly, we discuss the $H<0$ patch, where the universe begins with an initial finite-time singularity of Type III as $H$ and $\dot{H}$ diverge at $t\to t_s$, where
$$t_{s}=\int_{H_{0}<0}^{-\infty}\frac{dH}{\dot{H}}=\left(\frac{H_{0}}{\beta}\right)^{1/\alpha},$$
from some value $H_{0}<0$. Then, it evolves in the increasing $H$ direction towards a future fixed point $H=0$. However, the time requires to approach that point is infinite, so the universe has no future finite-time singularity. Secondly, in the $H>0$ patch, the universe has no initial singularity, since the time to reach its Minkowskian origin is infinite. However, the universe evolves in a phantom regime towards a future finite-time singularity of Type III as $t\to t_{s}$, where
$$t_{s}=\int_{H_{0}>0}^{\infty}\frac{dH}{\dot{H}}=\left(\frac{H_{0}}{\beta}\right)^{1/\alpha},$$
from some value $H_{0}>0$.\\

\textit{Case 2.} $\alpha=-\frac{1}{n}$ ($n>1$ is an even positive integer): The corresponding phase portraits are indicated by the dash curves as in Fig. \ref{Fig:phspTypeIII}\subref{fig:phsptypeIIIb}. Notably, in this subclass, the phase portraits corresponding to $\pm~ \beta$ are identical. In this case, the Minkowskian fixed point is stable (attractor), and therefore the $H<0$ and $H>0$ patches are separated by an infinite time. Therefore, we discuss each as a separate patch. At the positive $H$ patch, the cosmic behavior is the same as case 1. However, at the negative $H$ patch, the universe evolves effectively in phantom regime. It begins with a finite-time singularity at $t_{s}=\left(H_{0}/\beta\right)^{1/\alpha}$, then it evolves towards the right direction (increasing $H$). So the universe experiences an eternal accelerated contraction phase with a Minkowskian fate at an infinite time, i.e. it has no future singularity.  \\

\textit{Case 3.} $\alpha=-\frac{a}{b}$ ($a$ and $b$ are positives, and $a<b$ but $\frac{a}{b}\neq \frac{1}{n}$): The corresponding phase portraits are indicated by the dash curves as in Fig. \ref{Fig:phspTypeIII}\subref{fig:phsptypeIIIc}. Remarkably, the positive $H$ patch is not valid in this case. On the contrary, the dynamical behavior of the universe is the same as in cases 1 and 2 above, in the $H<0$ patch.
\subsubsection{$f_0>0$ ($\beta>0$)}\label{Sec:2.3.2}
In this category, similarly, three patterns could be obtained according to the value of $\alpha$. Therefore, we have the following subclasses:\\

\textit{Case 1.} $\alpha=-\frac{1}{n}$ ($n>1$ is an odd positive integer): The phase portraits for some particular choices of $\alpha$ are indicated by the solid curves as in Fig. \ref{Fig:phspTypeIII}\subref{fig:phsptypeIIIa}, where the Mikowskian fixed point at the origin of the phase space is a common point for all portraits. The Minkowski space, in this case, is a semi-stable fixed point. Since the universe requires an infinite time to reach this Minkowskian space, we find that the regions, $H>0$ and $H<0$, are two separate patches. Firstly, we discuss the $H>0$ patch, where the universe evolves in the decreasing $H$ direction. As shown by the plots, the universe has an initial finite-time singularity with a decelerated expansion behavior, where the age of the universe, for some value $H_{0}>0$, can be given by
$$t_{s}=\int_{\infty}^{H_{0}}\frac{dH}{\dot{H}}=\left(\frac{H_{0}}{\beta}\right)^{1/\alpha}.$$
However, for proper choice of the parameter $\alpha$ and $\beta$, the universe can evolve from deceleration to acceleration phase. We assume that the transition has been occurred  at some value $H_{de}>0$, then the phase portrait should cut the zero acceleration curve (i.e. $\dot{H}=-H^{2}$) at that value. Using Eq. (\ref{autonomous}), at the transition point, the two parameters can be related by
\begin{equation}\label{transition}
    \beta=H_{de}\left(-\frac{H_{de}}{\alpha}\right)^{\alpha}.
\end{equation}
For any value of $-1<\alpha<0$, the above equation can predict the value of $\beta$ when the Hubble parameter (or the red-shift) at transition is accurately measured by observations. For more detailed discussion about the possible values of $\alpha$ and $\beta$ to realize that transition, see Sec. \ref{Sec:2.5}. As shown in Fig. \ref{Fig:phspTypeIII}\subref{fig:phsptypeIIIa}, the universe evolves towards a quasi-de Sitter, i.e. $\omega_{\textmd{eff}}\rightarrow -1$, with a Minkowskian fate, which is unusual when dark energy interpreted as a cosmological constant. Secondly, we discuss the $H<0$ patch, where the universe evolves in the decreasing $H$ direction. As shown by the plots, the universe evolves from the Minkowskian fixed point, so it has no initial finite-time singularity. However, it begins with an accelerated contraction, then enters a later decelerated contraction with a future finite-time singularity at $t_{s}=-\left|\frac{H_{0}}{\beta}\right|^{1/\alpha}$ from some value $H_{0}<0$. Similarly, in this patch, the transition from an accelerated to a decelerated contraction can be realized. In this case, if we assume a negative cosmological constant in Eq. (\ref{autonomous}), we expect the phase portrait to shift downwards avoiding the Minkowskian fixed point, where a turnaround cosmology occurs.\\

\textit{Case 2.} $\alpha=-\frac{1}{n}$ ($n>1$ is an even positive integer): The corresponding phase portraits are given in Fig. \ref{Fig:phspTypeIII}\subref{fig:phsptypeIIIb}. As mentioned in Sec. \ref{Sec:2.3.1}, the phase portraits of $\pm~ \beta$ are identical. Therefore, we expect the same description as given before.\\

\textit{Case 3.} $\alpha=-\frac{a}{b}$ ($a$ and $b$ are positives, and $a<b$ but $\frac{a}{b}\neq \frac{1}{n}$): The corresponding phase portraits are indicated by the solid curves as in Fig. \ref{Fig:phspTypeIII}\subref{fig:phsptypeIIIc}. Remarkably, the negative $H$ patch is not valid in this case. On the contrary, the positive $H$ patch is valid, whereas the dynamical evolution is just as in the cases 1 and 2.
\subsubsection{the physical description}\label{Sec:2.3.3}
In conclusion, we find out the choice of $\beta>0$ provides an alternative to the $\Lambda$CDM models. In this scenario, the universe begins with an initial singularity of Type III, instead of Type I of the standard cosmology (big bang). Then, the universe traverses
from deceleration to acceleration. However, it evolves towards a Minkowskian fate not de Sitter, which is distinguishable from $\Lambda$CDM models when the dark energy is interpreted as a cosmological constant. In Sec. \ref{Sec:4.3}, we provide a complementary study through the torsion gravity to explain the late accelerating expansion phase of the model.
\subsection{Type IV singularity phase portrait}\label{Sec:2.4}
\begin{figure}
\centering
\includegraphics[scale=.29]{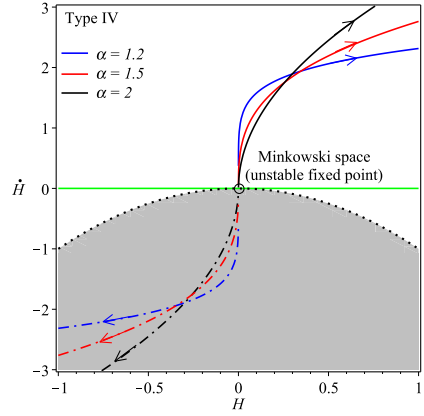}
\caption{The phase portraits of the finite-time singularities of Type IV for different choices of the parameter $\alpha>1$ in Eqs. (\ref{autonomous}). The solid curves are corresponding to $f_0>0$ ($\beta<0$), and the dash curves are corresponding to $f_0<0$ ($\beta>0$). Same color curves means we have used the same value of $\alpha$.}
\label{Fig:phspTypeIV}
\end{figure}

This singularity occurs when the cosmic time approaches $t \to t_s$, all the three quantities, $a$, $\rho_\mathrm{eff}$ and $\left|p_\mathrm{eff}\right|$, approach finite values, i.e. $a \to a_s$, $\rho_\mathrm{eff} \to \rho_s$ and $\left|p_\mathrm{eff}\right| \to p_s$. In addition, the Hubble parameter and its first derivative are finite, while its second/higher derivatives diverge. Type IV singularity is the softest (not of the crushing type) among the other four types, c.f. \cite{Nojiri:2005sx,Odintsov:2015ynk,Odintsov:2015gba,Odintsov:2015jca,Kleidis:2016vmd,Brevik:2016aos}. Using the scale factor (\ref{scale.factor}), the Type IV singularity case occurs when $\alpha > 1$. In this case, similar to Type I singularity, we have $\beta=f_0(1+\alpha)<0$, when $f_{0}>0$, while $\beta=f_0(1+\alpha)>0$, when $f_{0}<0$. Different cases of the phase portraits corresponding to Eq. (\ref{autonomous}) are given in Fig. \ref{Fig:phspTypeIV}. Although the singularity of Type IV is a fixed point as well, as appears in the phase portrait, it can be reached in a finite-time. This argument can be verified as follows. Since, in the Type IV case, the singularity occurs when the higher derivatives of the Hubble rate are divergent as $H \to H_{s}$, this means that
$$\lim_{H \to H_{s}} \frac{d^{n}H}{dt^{n}} = \pm \infty,$$
for some $n \geq 2$. Let us compute the lowest derivative for which the Type IV singularity could occur, which is for
$n = 2$,
$$\lim_{H \to H_{s}} \ddot{H} = \lim_{H \to H_{s}} \dot{H}\left(\frac{d\dot{H}}{dH}\right) = \pm \infty.$$
Since $\dot{H}$ is finite, it implies that
\begin{equation}\label{Type-IV}
\lim_{H \to H_{s}} \frac{d\dot{H}}{dH} = \pm \infty.
\end{equation}
This can be shown graphically on the phase space, Fig. \ref{Fig:phspTypeIV}, as an infinite slope of the phase portrait at the Type IV singularity phase point. Remarkably, this type of singularities is the only one which coincides with a fixed point among all other types. In general, the fixed points dominate the one dimensional autonomous systems, where the time required to reach these points is infinite as shown by Eq. (\ref{time:fixed-point}). However, Type IV singularities are exceptions and the system can reach them in a finite-time. In order to clarify this point, we rewrite the Friedmann equation (\ref{effective}) as
$$\dot H= -\frac{~\kappa^{2}}{2}\left(p_{\textmd{eff}}+\rho_{\textmd{eff}}\right).$$
It has been shown that for an equation of state $p_{\textmd{eff}}(H)$, and where the pressure is continuous and differentiable; the solution always reaches a fixed point in an infinite time \cite{Awad2013}. This left us with the other option where the pressure is not differentiable, i.e. $dp_{\textmd{eff}}/dH$ is not continuous. In this case, we, also, could have two possible cases: The first is when the discontinuity of $dp_{\textmd{eff}}/dH$ is finite, the time to reach a fixed point is also infinite. The second case when $dp_{\textmd{eff}}/dH$ is infinite discontinuous, it is the only possible option to reach a fixed point in a finite-time\footnote{In the general relativistic picture, condition (\ref{Type-IV}) leads to a divergence of the speed of sound $dp_{m}/d\rho_{m}=c_{s}^2$, so the solution will not be causal.}. Since $\rho_{\textmd{eff}} \sim H^{2}$, we see that $dp_{\textmd{eff}}/dH$ is infinite discontinuous if $d\dot{H}/dH$ diverges. Therefore, we write the following conditions for a fixed point $H_{f}$ to be reached in a finite time
\begin{itemize}
  \item [(i)] $\lim_{H\rightarrow H_{f}} \dot{H}=0$,
  \item [(ii)] $\lim_{H\rightarrow H_{f}} d\dot{H}/dH=\pm \infty$,
  \item [(iii)] $t=\int_{H}^{H_{f}}dH/\dot{H}<\infty$.
\end{itemize}
The above conditions are always fulfilled in the case of finite-time singularities of Type IV. Although the above calculations have been carried out for the lowest divergent derivatives $\ddot{H}\to \pm \infty$, it can be generalized for other higher lowest divergent derivatives of the Hubble parameter. We next turn our discussions of the phase portraits associated to the singularities of Type IV for some particular values of the model parameters $\alpha$ and $\beta$.
\subsubsection{$f_0>0$ ($\beta<0$)}\label{Sec:2.4.1}
The corresponding phase portraits are indicated by the solid curves as in Fig. \ref{Fig:phspTypeIV}. The universe has an initial singularity of Type IV at the Minkowskian unstable fixed point, and then it evolves in phantom regime. Although, the phase portrait evolves towards $H\to \infty$, it will not have a future finite-time singularity as clarified earlier after Eq. (\ref{autonomous}).
\subsubsection{$f_0<0$ ($\beta>0$)}\label{Sec:2.4.2}
The corresponding phase portraits are indicated by the dash curves as in Fig. \ref{Fig:phspTypeIV}. Similarly, the universe begins with a finite-time singularity of Type IV at a Minkowskian unstable fixed point, then it evolves in a non-phantom regime. It evolves towards $H\to -\infty$ with an early decelerated contraction phase. After, the universe enters a later accelerating contraction phase. Although the phase portrait evolves towards a big crunch singularity, it cannot be reached in a finite time as clarified before. Therefore, the universe will not have a future finite-time singularity.
\subsubsection{the physical description}\label{Sec:2.4.3}
In conclusion, we mention that the universe takes an infinite-time to leave or to reach a fixed point. So if the universe begins at a fixed point, it stays forever at that point. However, singularities of Type IV allow the universe to leave or to reach fixed points in a finite-time. So these types of singularities may provide an important key in inflationary models. In addition, if the phase portrait is a double valued about Type IV singularity, the universe is capable to cross the phantom divide line. The latter situation provides an important dynamical features in bouncing cosmology.\\

In the case at hand, which is shown in Fig. \ref{Fig:phspTypeIV}, the plots show that the cosmic fluid flow is towards the right direction (increasing $H$), whereas the Minkowskian fixed point coincides with the singular point of Type IV. The time to reach the initial singularity is $t=\left(\frac{H_{0}}{\beta}\right)^{1/\alpha}$ for a some value $H_{0}$. On the other hand, the universe has no future singularity as the time required to approach $H \to \infty$ is an infinite.
\subsection{Generic remarks}\label{Sec:2.5}
\begin{figure}
\centering
\subfigure[~Full range of four singularity types]{\label{fig:trans}\includegraphics[scale=.29]{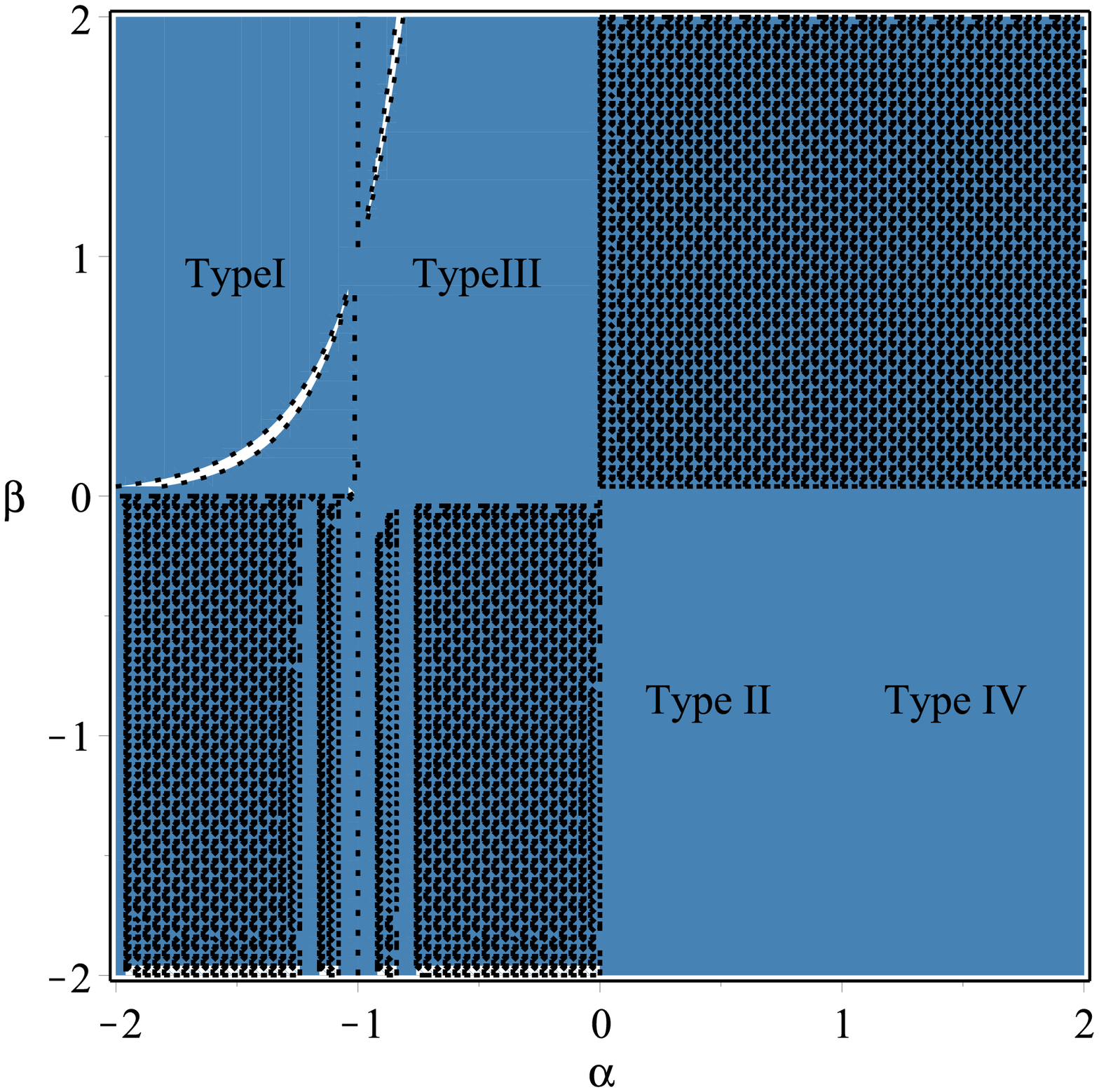}}
\subfigure[~Type I]{\label{fig:trans-typeI}\includegraphics[scale=.29]{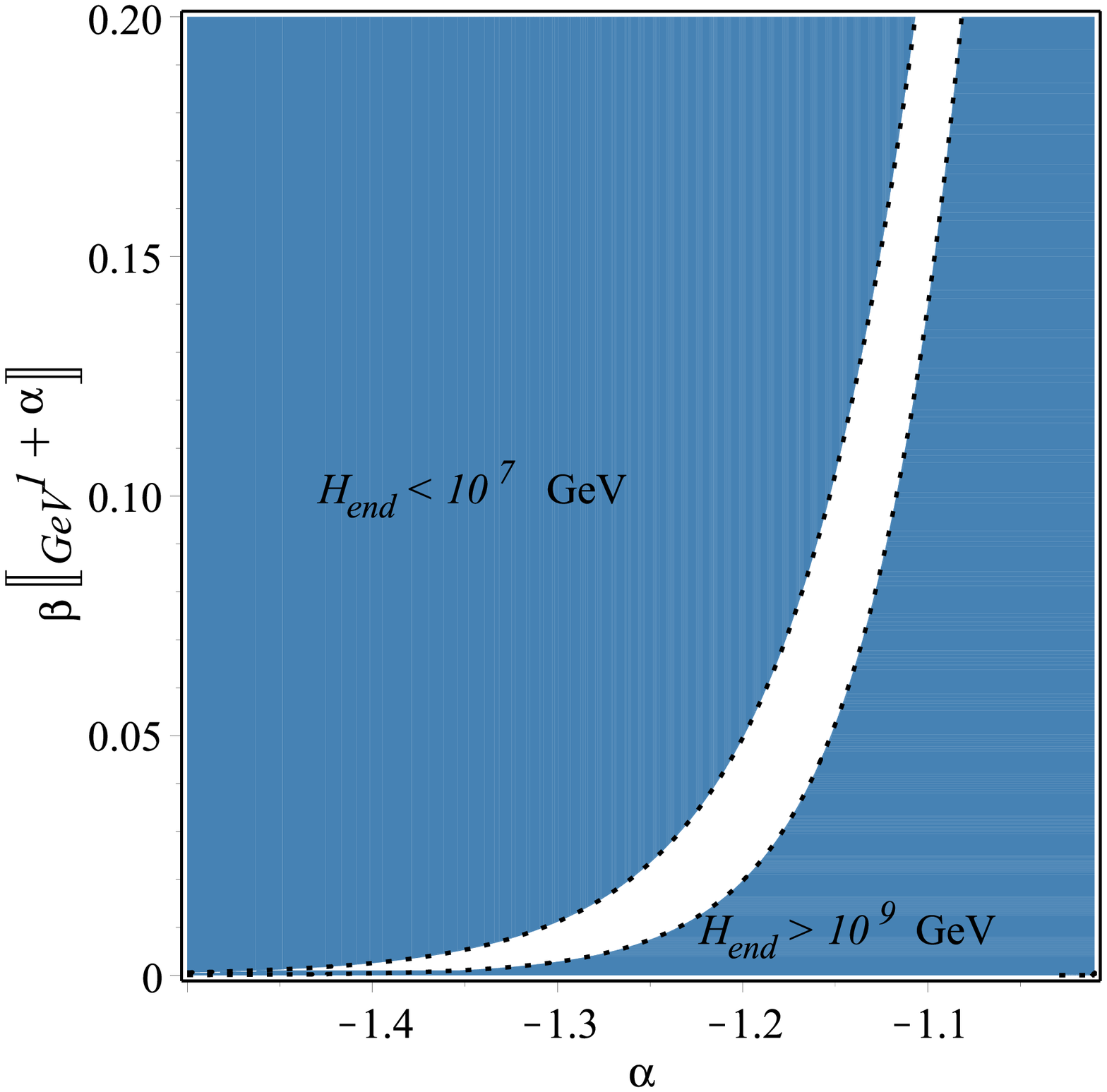}}
\subfigure[~Type III]{\label{fig:trans-typeIII}\includegraphics[scale=.29]{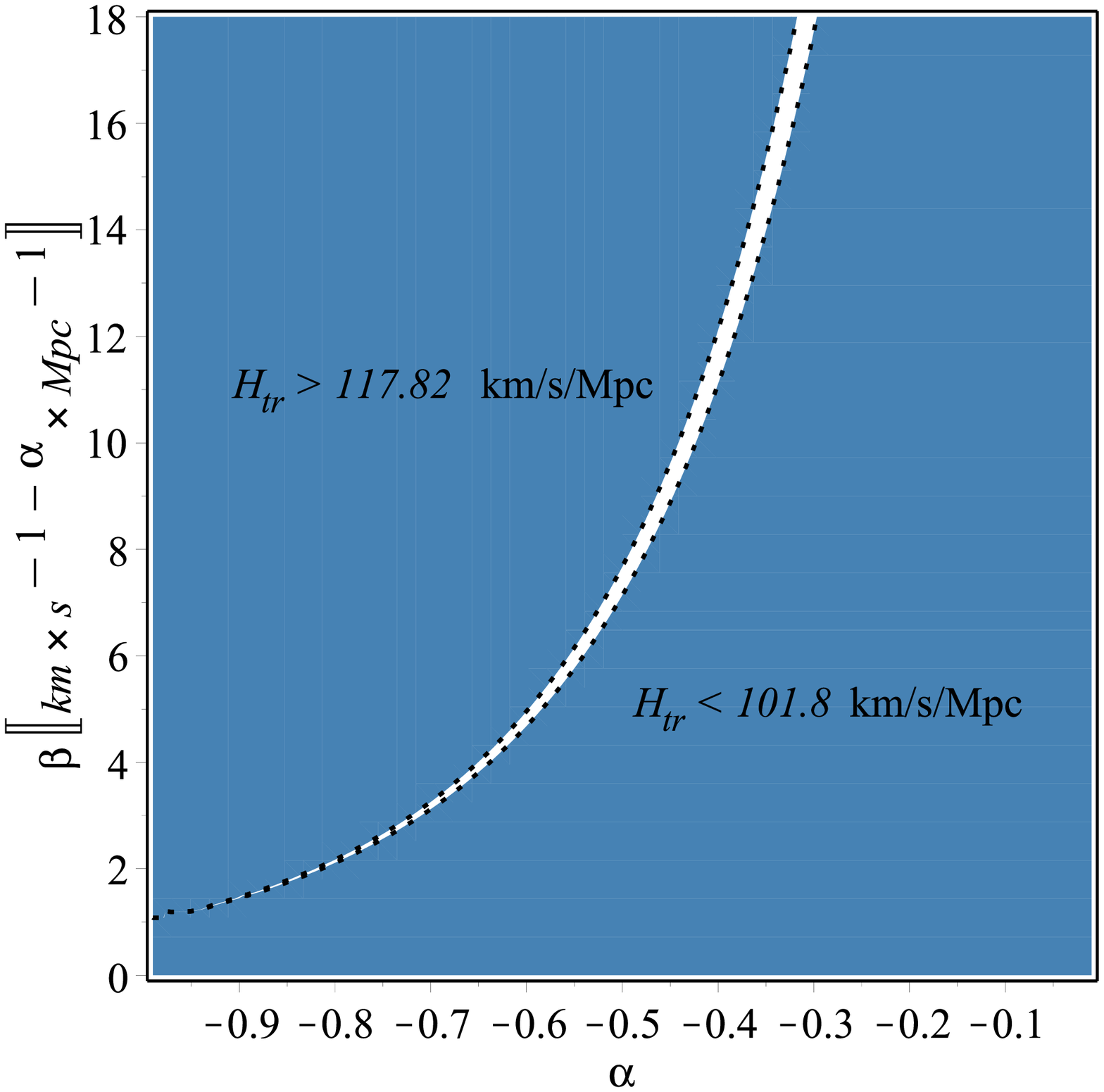}}
\caption[figtopcap]{\subref{fig:trans} ($\alpha$, $\beta$) diagram shows the values which allow transitions (\ref{inf-end}) and (\ref{transition}), where the dark regions are forbidden values, the blue regions are excluded values, while the white regions are the accepted values; \subref{fig:trans-typeI} values of the parameters $\alpha$ and $\beta$ allow transition from acceleration to deceleration at $10^7 \lesssim H_{inf} \lesssim 10^9$ GeV, and the dot curves give the values $\alpha$ and $\beta$ which allow transition at $H_{inf}=10^7$ and $10^9$ GeV exactly.; \subref{fig:trans-typeIII} values of the parameters $\alpha$ and $\beta$ which allow transition from deceleration to acceleration at redshift $0.72 \lesssim z_{de} \lesssim 0.84$ (i.e. $101.8 \lesssim H_{de} \lesssim 117.82$ km/s/Mpc), and the dot curves give the values $\alpha$ and $\beta$ which allow transition at $H_{de}=101.8$ and $117.82$ km/s/Mpc exactly.}
\label{Fig:alpha-beta}
\end{figure}

As shown in the previous sections, how phase portrait analysis is a qualitative powerful tool to extract dynamical information by fitting all possible universes on a small piece of paper. In this section, we summarize some general conclusions which characterize the dynamical evolution associated with each finite-time singularity type. In addition, we extract some useful quantitative information to conform the evolution to some important cosmic events. We noted that transition from acceleration to deceleration, see Sec. \ref{Sec:2.1}, or from deceleration to acceleration, see Sec. \ref{Sec:2.3}, could be realized for $H>0$ only in Type I or Type III singularities, respectively. This conclusion can be seen clearly using Fig. \ref{Fig:alpha-beta}\subref{fig:trans}.\\
Type I ($\alpha<-1$): For the values $f_{0}<0$ ($\beta>0$), the transition from acceleration to deceleration is allowed, see Fig. \ref{Fig:phspTypeI}\subref{fig:phsptypeIb}. So this model is suitable to describe a graceful exit inflationary models. Where the universe begins with a big bang inflation, then it exits into a decelerated FLRW phase. Since we deal with high energy scale, it is convenient to measure the Hubble parameter in the Planck unit. We assume that the inflation ends at $10^{-34}\lesssim t\lesssim 10^{-31}$ s, i.e. $10^{7}\lesssim H_{inf}\lesssim 10^{9}$ GeV. Substituting into Eq. (\ref{inf-end}), we obtain the values of ($\alpha$, $\beta$) which allow a viable graceful exit inflation. We represent the solution graphically in Fig. \ref{Fig:alpha-beta}\subref{fig:trans-typeI}.\\

Type II ($0<\alpha<1$): For particular choices of $\alpha=\frac{1}{n}$, where $n$ is a positive odd integer, we have a bouncing cosmology with a singularity of type II at the bouncing time, see Fig. \ref{Fig:phspTypeII}\subref{fig:phsptypeIIa}. Here junction conditions should be used to weld the contraction and the expansion phases. In order to avoid the trans-Planckian problem of inflationary model, it is convenient to choose the finite value $\rho_{m}$ to be at the Planckian density limit, or the minimal Hubble to have the Planck length. In this case we would have a non-singular bouncing scenario can avoid the trans-Planckian problems of the inflationary models.\\

Type III ($-1<\alpha<0$): We noted that the transition from deceleration to acceleration, in the $H>0$ region, can be achieved. So this model provides an alternative to $\Lambda$CDM universe. However, it evolves towards Minkowskian universe not de Sitter, which is unusual when dark energy is interpreted as a cosmological constant. In general, if the red-shift at transition (or the Hubble value $H_{de}$) is accurately measured by observations, we can solve Eq. (\ref{transition}) to obtain viable solutions of ($\alpha$, $\beta$). Since we deal with low energy scale, it is convenient to measure the Hubble parameter in SI units\footnote{Remember that 1 Mpc = 3.09 $\times$ 10$^{19}$ km, so the quantity km/Mpc is dimensionless. Consequently, the Hubble parameter is measured in [s]$^{-1}$.} [km/s/Mpc]. A recent analysis \cite{Farooq:2016zwm} shows that the deceleration to acceleration transition is at $z_{de}=0.72\pm 0.05$ ($0.84\pm 0.03$) for the present Hubble constant is taken as $H_{0}=68\pm 2.8$ ($73.24\pm1.74$) km/s/Mpc. We provide a graphical solution in Fig. \ref{Fig:alpha-beta} by assuming that the transition to be occurred in the range $101.8 < H_{de} < 117.82$ km/s/Mpc.\\

Type IV ($\alpha>1$): Remarkably, the singular point in this model is a repeller fixed point as well. Usually, the universe takes an infinite time to reach that point. However, in Type IV singularity the higher derivative of the Hubble parameter diverges. Consequently, we have $d\dot{H}/dH \to \pm \infty$. This provides a unique case, that is the fixed point can be reached in a finite time. So we may use this model to cross the phantom divide line between phantom and non-phantom regimes, if the phase portrait is a double valued function about the fixed point.\\

As shown in the above discussion that the phase space analysis is a useful tool to understand the evolution of the universe in a clear and transparent way. However, we also need to reformulate this description within a field theory framework for better understanding.
\section{Generalized teleparallel gravity}\label{Sec:3}
In the general relativity (GR) theory, the Friedmann system provides a power law scale factor, if we assume that the matter content is a perfect fluid with a linear (fixed) equation of state parameter $\omega_{m}$ such that $0\leq \omega_{m}\leq 1$ to grantee the stability and the causality conditions. Otherwise, one need to assume that the matter content to have an exotic equation of state, c.f. \cite{Awad2013,Nojiri:2015wsa,Brevik:2016aos}. In fact, the power law scale factor can predict perfectly the thermal history during radiation/matter dominant eras. However, it fails to describe the early and the late accelerating expansion epochs. In this sense, modified gravity may explain these epochs by modifying the gravitational sector.

On the contrary, as we clarified earlier that the phase portrait analysis through the stability of its fixed points requires that the Friedman equations to be written as a one dimensional autonomous system. Among several extensions of the GR we note that its teleparallel equivalent version (TEGR), c.f. \cite{M2013}, and its $f(T)$ extension are having a nice feature making them in comfort with our phase portrait analysis \cite{AHNS:2017}, that is the teleparallel torsion is proportional to the quadratic Hubble parameter so that the modified gravitational sector does not introduce the second or any higher derivatives of Hubble. For other extensions of the teleparallel gravity see \cite{KS114,Capozziello:2016eaz,Martin1,Martin2}. Since the modified Friedmann equations according to the $f(T)$ gravity contain only Hubble parameter and its rate of change, it is consistent with the phase space analysis. Therefore, the $f(T)$ theories, among many modified gravity theories, can be considered as a natural extension to the GR.
\subsection{Teleparallel space}\label{Sec:3.1}
In the Riemannian space of $n$-dimension, the metric tensor field $g_{\mu\nu}$ is the fundamental quantity. Instead, the vielbein vector field $e_{a}{^{\mu}}$ is the fundamental quantity in the Weitzenb\"{o}ck space of $n$-dimension\footnote{The Latin indices are called the Lorentz indices and the Greek indices are the coordinate indices, where both are running from $1,\cdots, n$ and additionally following the Einstein summation convention.}. The later can be defined as a pair $(M,e_{a})$, where $M$ is an $n$-dimensional differentiable manifold and the set $\{e_{a}\}$ contains $n$ independent vector fields defined globally on $M$, this set at point $p$ is the basis of its tangent space $T_{p}M$.  Because of the independence of $e_{a}$, the determinant $e\equiv \det (e_{a}{^{\mu}})$ is nonzero. Then, one can define the linear (Weitzenb\"{o}ck) connection
\begin{equation}\label{W_connection}
\Gamma^{\alpha}{_{\mu\nu}}\equiv e_{a}{^{\alpha}}\partial_{\nu}e^{a}{_{\mu}}=-e^{a}{_{\mu}}\partial_{\nu}e_{a}{^{\alpha}}.
\end{equation}
This connection is characterized by the property that
\begin{equation}\label{AP_condition}
\nabla_{\nu}e_{a}{^{\mu}}\equiv\partial_{\nu}{e_a}^\mu+{\Gamma^\mu}_{\lambda \nu} {e_a}^\lambda\equiv 0,
\end{equation}
where the covariant derivative $\nabla_{\nu}$ is associated to the Weitzenb\"{o}ck connection. This nonsymmetric connection uniquely determines the teleparallel geometry, since the vielbein vector fields are parallel with respect to it. Indeed, a vielbein space admits the dual and the symmetric connections associated to the Weitzenb\"{o}ck except they do not provide a teleparallel geometry. For more details about these connections and their applications see \cite{MW77,MW81,W85,W86,W98,MWE95,W99,NA07,NA08,NW13}. Also, for more Refs. on the parameterized versions of these connections and their applications see \cite{W2000,W2001,WMK2000,WMK2001,W2007,W2012,WS2010,Wanas2016,Wanas:2014lda}. However, the vielbein vector fields satisfy
\begin{equation}\label{orthonormality}
e_{a}{^{\mu}}e^{a}{_{\nu}}=\delta^{\mu}_{\nu}\quad \textmd{and}\quad e_{a}{^{\mu}}e^{b}{_{\mu}}=\delta^{b}_{a},
\end{equation}
where $\delta$ is the Kronecker tensor. Thus, we can construct an associated (pseudo-Riemannian) metric for any set of basis
\begin{equation}\label{metric}
g_{\mu \nu} \equiv \eta_{ab}e^{a}{_{\mu}}e^{b}{_{\nu}},
\end{equation}
while the inverse metric
\begin{equation}\label{inverse}
g^{\mu \nu} = \eta^{ab}e_{a}{^{\mu}}e_{b}{^{\nu}}.
\end{equation}
Also, it can be shown that $e=\sqrt{-g}$, where $g\equiv \det(g)$. In this sense, the vielbein space is a pseudo-Riemannian as well. Thus, we go further to define the symmetric Levi-Civita connection
\begin{equation}\label{Christoffel}
\overcirc{\Gamma}{^{\alpha}}{_{\mu\nu}}= \frac{1}{2} g^{\alpha \sigma}\left(\partial_{\nu}g_{\mu \sigma}+\partial_{\mu}g_{\nu \sigma}-\partial_{\sigma}g_{\mu \nu}\right).
\end{equation}
Recalling the absolute parallelism condition (\ref{AP_condition}), it is easy to show that the Weitzenb\"{o}ck and the Levi-Civita connections are both metric connections, i.e.
$$\nabla_{\sigma}g_{\mu\nu}\equiv 0,\quad \overcirc{\nabla}_{\sigma}g_{\mu\nu}\equiv 0,$$
where $\overcirc{\nabla}_{\nu}$ is the covariant derivative associated to the Levi-Civita connection.

The noncommutation of an arbitrary vector fields $V_{a}$ is given by
\begin{eqnarray}
\nonumber \nabla_{\nu}\nabla_{\mu}V_{a}{^{\alpha}} - \nabla_{\mu}\nabla_{\nu}V_{a}{^{\alpha}} &=& R^{\alpha}{_{\epsilon\mu\nu}}
V_{a}{^{\epsilon}} + T^{\epsilon}{_{\nu\mu}} \nabla_{\epsilon} V_{a}{^{\alpha}},\\
\nonumber \overcirc{\nabla}_{\nu}\overcirc{\nabla}_{\mu}V_{a}{^{\alpha}} - \overcirc{\nabla}_{\mu}\overcirc{\nabla}_{\nu}V_{a}{^{\alpha}} &=& \overcirc{R}{^{\alpha}}{_{\epsilon\mu\nu}}
V_{a}{^{\epsilon}} + \overcirc{T}{^{\epsilon}}{_{\nu\mu}} \overcirc{\nabla}_{\epsilon} V_{a}{^{\alpha}},
\end{eqnarray}
where $R^{\alpha}{_{\epsilon\mu\nu}}$ ($\overcirc{R}{^{\alpha}}{_{\epsilon\mu\nu}}$) and $T^{\epsilon}{_{\nu\mu}}$ ($\overcirc{T}{^{\epsilon}}{_{\nu\mu}}$) are the curvature and the torsion tensors of the weitzenb\"{o}ck (Levi-Civita) connection, respectively. The absolute parallelism condition (\ref{AP_condition}) and the noncommutation formula force the curvature tensor $R^{\alpha}_{~~\mu\nu\sigma}$ of the Weitzenb\"{o}ck connection to vanish identically, i.e. $R^{\alpha}{_{\epsilon\mu\nu}}\equiv 0$, while the symmetric Levi-Civita connection provides a vanishing torsion tensor, i.e. $\overcirc{T}{^{\epsilon}}{_{\nu\mu}} \equiv 0$.

The torsion tensor of the Weitzenb\"{o}ck connection (\ref{W_connection}) is defined as
\begin{equation}
T^\alpha{_{\mu\nu}}\equiv{\Gamma^\alpha}_{\nu\mu}-{\Gamma^\alpha}_{\mu\nu}={e_a}^\alpha\left(\partial_\mu{e^a}_\nu
-\partial_\nu{e^a}_\mu\right).\label{Torsion}\\
\end{equation}
Then, the contortion tensor $K^{\alpha}_{~\mu\nu}$ is defined by
\begin{equation}
K^{\alpha}{_{\mu\nu}} \equiv \Gamma^{\alpha}_{~\mu\nu} - \overcirc{\Gamma}{^{\alpha}}_{\mu\nu}=e_{a}{^{\alpha}}~ \overcirc{\nabla}_{\nu}e^{a}{_{\mu}}. \label{contortion}
\end{equation}
It is to be noted that $T_{\mu\nu\sigma}$ is skew symmetric in the last pair of indices whereas $K_{\mu\nu\sigma}$ is skew symmetric in the first pair of indices. Moreover, the torsion and the contortion can be interchangeably following the useful relations:
\begin{equation}\label{torsion03}
        T_{\alpha \mu \nu}=K_{\alpha \mu \nu}-K_{\alpha \nu \mu},
\end{equation}
\begin{equation}\label{contortion03}
        K_{\alpha \mu \nu}=\frac{1}{2}\left(T_{\nu\alpha\mu}+T_{\alpha\mu\nu}-T_{\mu\alpha\nu}\right).
\end{equation}

In the teleparallel space there are three Weitzenb\"{o}ck invariants: $I_{1}=T^{\alpha \mu \nu}T_{\alpha \mu \nu}$, $I_{2}=T^{\alpha \mu \nu}T_{\mu \alpha \nu}$ and $I_{3}=T^{\alpha}T_{\alpha}$, where $T^{\alpha}=T_{\rho}{^{\alpha \rho}}$. We next define the invariant $$T=\frac{1}{4} I_{1}+\frac{1}{2} I_{2} - I_{3},$$
by combining the three invariants $I_1$, $I_2$ and $I_3$ with the prefixes coefficients as appears above. This teleparallel invariant is equivalent to the Ricci scalar $\overcirc{R}$ up to a total derivative term as we will show below. Alternatively, the teleparallel torsion scalar is given in the compact form
\begin{equation}
T \equiv {T^\alpha}_{\mu \nu}{S_\alpha}^{\mu \nu},\label{Tor_sc}
\end{equation}
where the superpotential tensor
\begin{equation}
{S_\alpha}^{\mu\nu}=\frac{1}{2}\left({K^{\mu\nu}}_\alpha+\delta^\mu_\alpha{T^{\beta\nu}}_\beta-\delta^\nu_\alpha{T^{\beta \mu}}_\beta\right),\label{superpotential}
\end{equation}
is skew symmetric in the last pair of indices. Indeed, we deal with exactly one space. However, Levi-Civita and Weitzenb\"{o}ck connections can see this space with different resolutions. The former represents an extreme picture with a vanishing torsion tensor, while the later represents another extreme with a vanishing curvature tensor. Interestingly, one can find possible links between these two extremes. Thus, a useful link in this context is the following identity
\begin{equation}\label{identity}
    e\overcirc{R} \equiv -e T + 2\partial_{\mu}(e T^{\mu}),
\end{equation}
where divergence term sometimes given in terms of the Levi-Civita covariant derivative as $\partial_{\mu}(e T^{\mu})=e \overcirc{\nabla}_{\mu} T^{\mu}$. Since the Ricci and the teleparallel torsion scalars differ by a total derivative term, both would provide the same set of field equations using the lagrangian formalism, GR and TEGR, respectively. Although the superficial equivalence on the level of the field equations, the difference runs deep on the lagrangian level. It can be shown that the left hand side of the geometrical identity (\ref{identity}) is a diffeomorphism scalar and Lorentz as well. However, the total derivative term in the right hand side is not a Lorentz scalar. Consequently, the teleparallel torsion scalar density is a diffeomorhism scalar but not a Lorentz scalar. This contrast is crucial in the generalization of the lagrangian of the GR and the TEGR theories by taking, respectively, $f(\overcirc{R})$ and $f(T)$ extensions. The former provides field equations invariant under local Lorentz transformation, while the later is not \cite{1010.1041,1012.4039,LLT1,Nashed:2009hn}. We consider the action of the $f(T)$ gravity \cite{BF09,L10}
\begin{equation}\label{action}
{\cal S}=\frac{1}{2 \kappa^2}\int d^{4}x~  e f(T)+{\cal S}_{m},
\end{equation}
where $\cal{S}_{m}$ is the matter action. The variation of the action (\ref{action}) with respect to the tetrad gives
\begin{equation}\label{field_eqns}
\frac{1}{e} \partial_\mu \left( e S_a^{\verb| |\mu\nu} \right) f^{\prime}-e_a^\lambda  T^\rho_{\verb| |\mu \lambda} S_\rho^{\verb| |\nu\mu}f^{\prime}
+S_a^{\verb| |\mu\nu} \partial_\mu T f^{\prime\prime}
+\frac{1}{4} e_a^\nu f=\frac{{\kappa}^2}{2} e_a^\rho \mathfrak{T}_\rho^{\verb| |\nu},
\end{equation}
where $f=f(T)$, $f^{\prime}=\frac{\partial f(T)}{\partial T}$, $f^{\prime\prime}=\frac{\partial^2 f(T)}{\partial T^2}$, and ${\mathfrak{T}_{\mu}}^{\nu}$ is the usual energy-momentum tensor of matter fields. It is clear that the field equations (\ref{field_eqns}) reproduce the TEGR theory by setting $f(T)=T$. The $f(T)$ modified gravity theories have been used widely in literature in cosmology \cite{Capozziello:2011hj,Cai:2011tc,Saridakis2,Bamba:2011pz,Mohseni:2012PhLB, Bamba:2012vg,Bamba:2013ooa,bounce3,BGLL2011,Bamba2016,Nashed20141,Nd15,ElHanafy2015,ElHanafy2016a,ElHanafy2016c}, and in astrophysical applications \cite{Iorio,ElHanafy2016b,DeBenedictis:2016aze,Farrugia:2016xcw,Junior:2015fya,Capozziello:2012zj,
Nashed2017,Nashed:2003ee,Nashed:2013bfa,Nashed:uja,2012ChPhL..29e0402G}, for more details about the $f(T)$ gravity see the review \cite{Saridakis1}.
\subsection{Reconstructing $f(T)$}\label{Sec:3.2}
We consider the diagonal vierbein corresponding to the FLRW metric (\ref{FLRW-metric}), i.e.
\begin{equation}\label{tetrad}
{e_{\mu}}^{a}=\textmd{diag}(1,a(t),a(t),a(t)).
\end{equation}
This directly relates the teleparallel torsion scalar (\ref{Tor_sc}) to Hubble parameter as
\begin{equation}\label{TorHubble}
T=-6H^2,
\end{equation}
The useful relation above facilitates many cosmological applications in the $f(T)$ gravity. We assume that the stress-energy tensor to be for perfect fluid as
\begin{equation}\label{matter}
\mathfrak{T}_{\mu\nu}=\rho_{m} u_{\mu}u_{\nu}+p_{m}(u_{\mu}u_{\nu}-g_{\mu\nu}),
\end{equation}
where $u^{\mu}$ is the fluid 4 velocity, $\rho_{m}$ and $p_{m}$ are the energy density and pressure of the fluid in its rest frame. Inserting the vierbein (\ref{tetrad}) into the field equations (\ref{field_eqns}) for the matter fluid (\ref{matter}), the modified Friedmann equations of the $f(T)$-gravity read
\begin{eqnarray}
  \rho_{m} &=& \frac{1}{2\kappa^2}\left[f(T)+12 H^2 f_{T}\right], \label{FR1T}\\
  p_{m} &=& \frac{-1}{2\kappa^2}\left[f(T)+4(3H^2+\dot{H})f_{T}-48\dot{H}H^2 f_{TT}\right].\label{FR2T}
\end{eqnarray}
In the above, the usual Friedmann equations are recovered by setting $f(T)=T$. Assuming that the matter fluid is governed by the linear equation of state $p_{m}=\omega_{m}\rho_{m}$, where $\omega_{m}=0$ for dust and $\omega_{m}=1/3$ for radiation, the system acquires the conservation (continuity) equation
\begin{equation}\label{continuity}
    \dot{\rho}_{m}+3H(1+\omega_{m})\rho_{m}=0.
\end{equation}
As mentioned before, the modified Friedmann equations of any $f(T)$-theory can be viewed as a \textit{one dimensional autonomous system}, i.e. $\dot{H}=\mathcal{F}(H)$, if we use the linear equation of state of the universe matter. So it is convenient now to represent Eqs. (\ref{FR1T}) and (\ref{FR2T}) in terms of $H$ \cite{AHNS:2017},
\begin{eqnarray}
  \rho_{m} &=& \frac{1}{2\kappa^2}\left[f(H)-H f_{H}\right], \label{FR1H}\\
  p_{m} &=& \frac{-1}{2\kappa^2}\left[f(H)-H f_{H}-\frac{1}{3}\dot{H} f_{HH}\right],\label{FR2H}
\end{eqnarray}
where $f_{H}:=\frac{df}{dH}$ and $f_{HH}:=\frac{d^{2}f}{dH^2}$. After some manipulation, we write
\begin{equation}\label{phasetrajectory}
    \dot{H}=3(1+\omega_{m})\left[\frac{f(H)-H f_{H}}{f_{HH}}\right]=\mathcal{F}(H).
\end{equation}
Combining (\ref{autonomous}) and (\ref{phasetrajectory}), we can obtain the $f(H)$ which produces some desired phase trajectory. The integral of the continuity equation (\ref{continuity}) can be given by
\begin{equation}\label{cont-Int}
    \rho_{m}=\rho_{m,0}~ e^{-3(1+\omega_{m})\displaystyle{\int} \tfrac{H}{\dot{H}}dH},
\end{equation}
where the integration constant
$$\rho_{m,0}\equiv \rho_{m}(t_{0}) \approx 1.88\times 10^{-26}~\Omega_{m,0}~ h_{0}^{2}~\textmd{kg/m}^{3},$$
the matter density parameter $\Omega_{m,0}$, and the dimensionless hubble constant $h_{0}$ are given by the observations at present time $t_{0}$. Substituting (\ref{autonomous}) into (\ref{cont-Int}), then Eq. (\ref{FR1H}) reads
\begin{equation}
\nonumber    f(H)-Hf_{H}=2\kappa^{2}\rho_{m,0}~ e^{-\tfrac{3(1+\omega_{m})H}{-1+\alpha}\left(\tfrac{H}{\beta}\right)^{-1/\alpha}}.
\end{equation}
Solving the above equation with respect to $f(H)$, we obtain
\begin{eqnarray}\label{f(H)}
\nonumber f(H)&=&A\left[B~\textmd{WhittakerM}\left(\frac{-2\alpha-1}{2+2\alpha},~ \frac{2+\alpha}{2+2\alpha}, \frac{-C}{1+\alpha}\right)\right.\\
& &\quad+\left.\frac{1}{3}~\textmd{WhittakerM}\left(\frac{1}{2+2\alpha},~ \frac{2+\alpha}{2+2\alpha}, \frac{-C}{1+\alpha}\right)\right],\qquad
\end{eqnarray}
where $A$, $B$ and $C$ are functions of the Hubble parameter, which can be listed as follows
\begin{eqnarray}
  A \equiv A(H) &=& \frac{6\rho_{m,0} \kappa^2}{2+\alpha}\left[\frac{1+\alpha}{3(1+\omega_{m})}\right]^{\tfrac{3+2\alpha}{2+2\alpha}}H^{-\tfrac{3+2\alpha}{2\alpha}}
  \beta^{\tfrac{3+2\alpha}{2\alpha(1+\alpha)}}e^{\tfrac{C}{2+\alpha}},\qquad \\[3pt]
  B \equiv B(H) &=& \frac{1+\alpha}{3}\left(1-C\right), \\[3pt]
  C \equiv C(H) &=& -3 \beta (1+\omega_{m}) H \left(\frac{H}{\beta}\right)^{\frac{1}{\alpha}}.
\end{eqnarray}
We note that in Eq. (\ref{f(H)}), we omitted a term $H \propto \sqrt{-T}$, because it has no contribution in the field equations. So we omit this term without loosing the generality of the solution. Substituting (\ref{f(H)}) into (\ref{FR1H}) and (\ref{FR2H}), the density and the pressure of the matter fluid read
\begin{eqnarray}
    \rho_{m}(H)&=&\rho_{m,0} e^{-\tfrac{3(1+\omega_m) H}{1+\alpha}\left(\tfrac{H}{\beta}\right)^{\tfrac{1}{\alpha}}},\label{matter-dens}\\
    p_{m}(H)&=&\omega_m \rho_{m,0} e^{-\tfrac{3(1+\omega_m) H}{1+\alpha}\left(\tfrac{H}{\beta}\right)^{\tfrac{1}{\alpha}}}.\label{matter-press}
\end{eqnarray}
By evaluating the density and the pressure of the ordinary matter, we can provide a complementary description of the cosmic evolution by identifying the gravitational sector contribution, in our case it is the torsion contribution.
\section{The Torsion Role}\label{Sec:4}
In order to manifest the torsion role in the cosmic evolution near the singularities, it is convenient to transform Eqs. (\ref{FR1H}) and (\ref{FR2H}) from the \textit{matter} frame to the \textit{effective} frame at which the equations would have the standard Einstein's field equations in addition to the torsion contribution as a higher order gravity of the $f(T)$ theory. So we write the modified Friedmann equations in the case of $f(T)$ gravity, i.e.
\begin{eqnarray}
{H}^2& =& \frac{\kappa^2}{3} \left( \rho_{m}+  \rho_{ T} \right)~\equiv \frac{\kappa^2}{3} \rho_{\textmd{eff}}, \label{MFR1}\\
2 \dot{{H}} + 3{H}^2&=& - \kappa^2 \left(p_{m}+p_{ T }\right)\equiv -\kappa^2 p_{\textmd{eff}},\label{MFR2}
\end{eqnarray}
where $\rho_{ T}$ and $p_{ T}$ are the effective density and pressure of the torsion fluid, respectively. By comparison with (\ref{FR1H}) and (\ref{FR2H}) we write
\begin{eqnarray}
\nonumber  \rho_{T}(H)&=&\frac{1}{2\kappa^2}[H f_{H}-f(H)+6H^{2}],\\
               &=&\frac{3}{\kappa^2}H^2-\rho_{m,0} e^{-\tfrac{3(1+\omega_m) H}{1+\alpha}\left(\tfrac{H}{\beta}\right)^{\tfrac{1}{\alpha}}}.\label{rhoT}\\
\nonumber    p_{T}(H)&=&-\frac{1}{6\kappa^2}\dot{H}\left(12+f_{HH}\right)-\rho_{T}(H),\\
\nonumber    &=&-\frac{1}{\kappa^2}\left[3H^2+2\alpha H \left(\frac{\beta}{H}\right)^{\tfrac{1}{\alpha}}\right]-\omega_m \rho_{m,0} e^{-\tfrac{3(1+\omega_m) H}{1+\alpha}\left(\tfrac{H}{\beta}\right)^{\tfrac{1}{\alpha}}}.\\\label{pT}
\end{eqnarray}
In the above we have replaced the value of $\dot{H}$ from (\ref{phasetrajectory}). One can show that $\rho_{T}$ and $p_{T}$ vanish where $f(H)=-6H^2$ and the standard Friedmann equations are recovered. In this case the effective torsion gravity acquires the conservation equation
\begin{equation}\label{continuityT}
    \dot{\rho}_{T}+3H\left[1+\omega_{T}(H)\right]\rho_{T}\equiv 0.
\end{equation}
We first evaluate the effective equation of state parameter, using Eq. (\ref{autonomous}), we obtain
\begin{eqnarray}\label{eff_EoS}
\nonumber \omega_{\textmd{eff}}&\equiv& \frac{p_{m}+p_{T}}{\rho_{m}+\rho_{T}}=-1-\frac{2\dot{H}}{3H^{2}},\\
&=&-1-\frac{2}{3}\frac{\alpha}{H} \left(\frac{\beta}{H}\right)^{\tfrac{1}{\alpha}}.
\end{eqnarray}
It is convenient to study the asymptotic behavior at large Hubble regimes where $H\to \pm \infty$. In general, for the power-law phase portraits $\dot{H}\propto H^{\gamma}$, we have $\omega_{\textmd{eff}} \to -1$ where $\gamma<2$, while $\omega_\textmd{eff}\to \pm \infty$ where $\gamma>2$. Also, we investigate the behavior of the effective equation of state at the Minkowskian fixed point $H=0$. We find that $\omega_\textmd{eff}\to \pm \infty$ where $\gamma<2$, while $\omega_\textmd{eff}\to -1$ where $\gamma>2$. For the model at hand, by recalling Eq. (\ref{eff_EoS}), at the limit $H\to \pm \infty$, we have the case of $\omega_\textmd{eff}\to -1$ where $\alpha<-1$ or $\alpha>0$, which covers the singularities of Types I, II and IV. For the same range, $\alpha<-1$ or $\alpha>0$, at the limit $H\to 0$, we obtain the following
\begin{equation}\label{Q0}
\lim_{H\to 0} \omega_\textmd{eff}=Q_{0},\quad \lim_{H\to \pm \infty} \omega_\textmd{eff}=-1,
\end{equation}
where $Q_{0}\equiv Q_{0}(H)=-\frac{2}{3}\frac{\alpha}{H}\left(\frac{\beta}{H}\right)^{1/\alpha}$. On the contrary, for the range $-1< \alpha < 0$, it is easy to verify that the effective equation of state having the following limits
\begin{equation}\label{Q02}
\lim_{H\to 0} \omega_\textmd{eff}=-1,\quad \lim_{H\to \pm \infty} \omega_\textmd{eff}=Q_{0},
\end{equation}
which describes the cosmic evolution associated to the Type III singularity. In conclusion, the effective equation of state evolves towards either $-1$ or $\pm \infty$, we summarize the behavior of the effective fluid in Table \ref{Table:EoS} in Sec. \ref{Sec:5}.\\

We next turn the discussion to the dynamical description instead of the above kinematical one. Therefore, we investigate the torsion role near the singularities using the torsion equation of state parameter, that is
\begin{eqnarray}\label{Tor_EoS}
\nonumber \omega_{T}&\equiv& \frac{p_{ T}}{\rho_{ T}}=-1+\frac{1}{3}\frac{\dot{H}\left(12+f_{HH}\right)}{f(H)-6H^{2}-H f_{H}},\\[3pt]
&=&-1+\frac{2\alpha H (\beta/H)^{\tfrac{1}{\alpha}}e^{\tfrac{3 (1+\omega_{m})H}{(1+\alpha)}\left(\tfrac{H}{\beta}\right)^{\tfrac{1}{\alpha}}}-(1+\omega_{m})\rho_{m,0}\kappa^2}{\rho_{m,0} \kappa^2-3 H^2 e^{\tfrac{3 (1+\omega_{m})H}{(1+\alpha)}\left(\tfrac{H}{\beta}\right)^{\tfrac{1}{\alpha}}}}.\qquad
\end{eqnarray}
The above expression has been evaluated by recalling Eqs. (\ref{rhoT}) and (\ref{pT}). In the physical models, we have $\kappa^2\equiv 8\pi G=1.68\times 10^{-9}$ m$^3$/kg/s$^{2}$, and $\rho_{m,0}=2.84\times 10^{-27}$ kg/m$^{3}$. Therefore, the value $\kappa^2 \rho_{m,0}\to 0$ s$^{-2}$ so that Eq. (\ref{Tor_EoS}) reduces to
$$\omega_{T}\to -1-\frac{2}{3}\frac{\alpha}{H} \left(\frac{\beta}{H}\right)^{\tfrac{1}{\alpha}}=\omega_{\textmd{eff}},$$
which coincides with the effective equation of state (\ref{eff_EoS}). In order to investigate the exact role of the torsion fluid, it is convenient to study the behavior of the torsion equation of state near the singularity. As clear from Eq. (\ref{Tor_EoS}) that the torsion equation of state is too sensitive to the choices of the parameters $\alpha$ and $\beta$. Therefore, we will discuss each case individually in the following sections.

In Sec. \ref{Sec:2}, we presented a generic study of the different types of the finite-time singularities by analyzing their phase portraits. In the following, we perform a complementary analysis of the cosmologies related to these singularity types through the effective and the torsion equations of state to find out the role of the torsion fluid and its behavior near these singularities. In fact, the cosmic accelerated expansion can be modeled using the cosmological constant in the simplest case, when this constant is introduced into Einstein's field equations as a matter source of the so called dark energy with a fixed equation of state $\omega=-1$. Although, it fits with Planck observations, it does not provide information about the nature of the dark energy. Other proposals have been introduced using a quintessence ($-1<\omega<-1/3$) or phantom ($\omega<-1$) scalar field, where crossing between these two regimes within the single canonical scalar field models is impossible \cite{Xia:2007km}. This is on the contrary of the quintom models, where crossing the phantom divide line can be achieved. A comprehensive criticism of quintom bounce is given in the review \cite{Cai:2009zp}. Usually, quintom models are realizable by introducing two scalar fields (quintessence + phantom) \cite{Feng:2004ad,Guo:2004fq}, or adding extra degrees of freedom by including higher derivative terms into the action \cite{Arefeva:2005mka,Li:2005fm,Arefeva:2006rnj}. As we have shown that the $f(T)$ cosmology provides a dynamical system equivalent to the general relativistic model when using an exotic equation of state, e.g. quintessence, phantom and quintom. This equivalence can be justified also by studying the role of the torsion equation os state.
\subsection{Type I singularity}\label{Sec:4.1}
Using Eqs. (\ref{eff_EoS}) and (\ref{Tor_EoS}), we plot their evolutions as given in Fig. \ref{Fig:EoS-TypeI}. As we have shown earlier in Sec. \ref{Sec:2.1}, there are two cases can be discussed for the models which have singularity of Type I ($\alpha<-1$):
\begin{figure}
\centering
\subfigure[~$f_0>0$ ($\beta<0$)]{\label{fig:EoS-typeIa}\includegraphics[scale=.3]{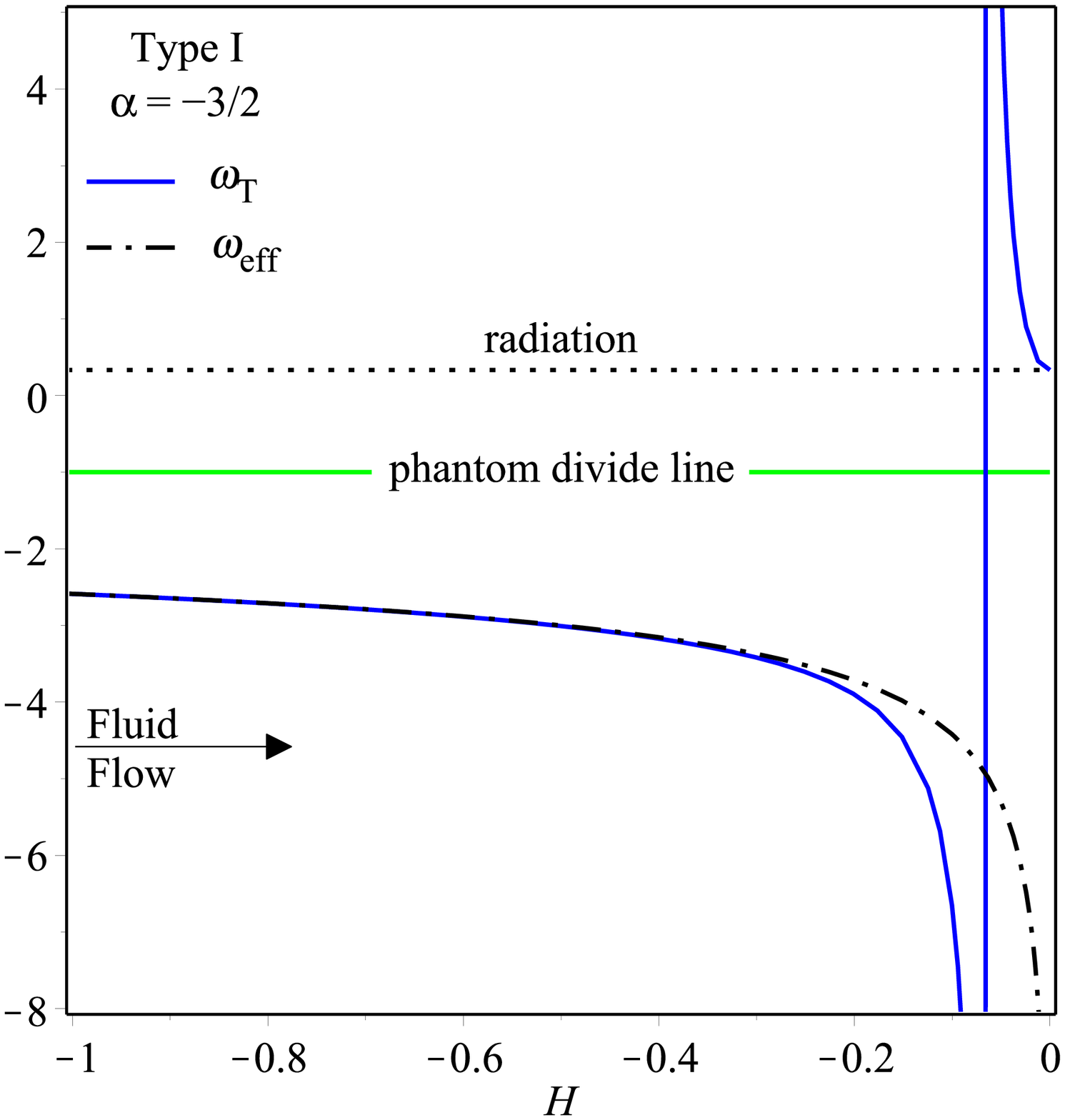}}\hspace{2mm}
\subfigure[~$f_0<0$ ($\beta>0$)]{\label{fig:EoS-typeIb}\includegraphics[scale=.3]{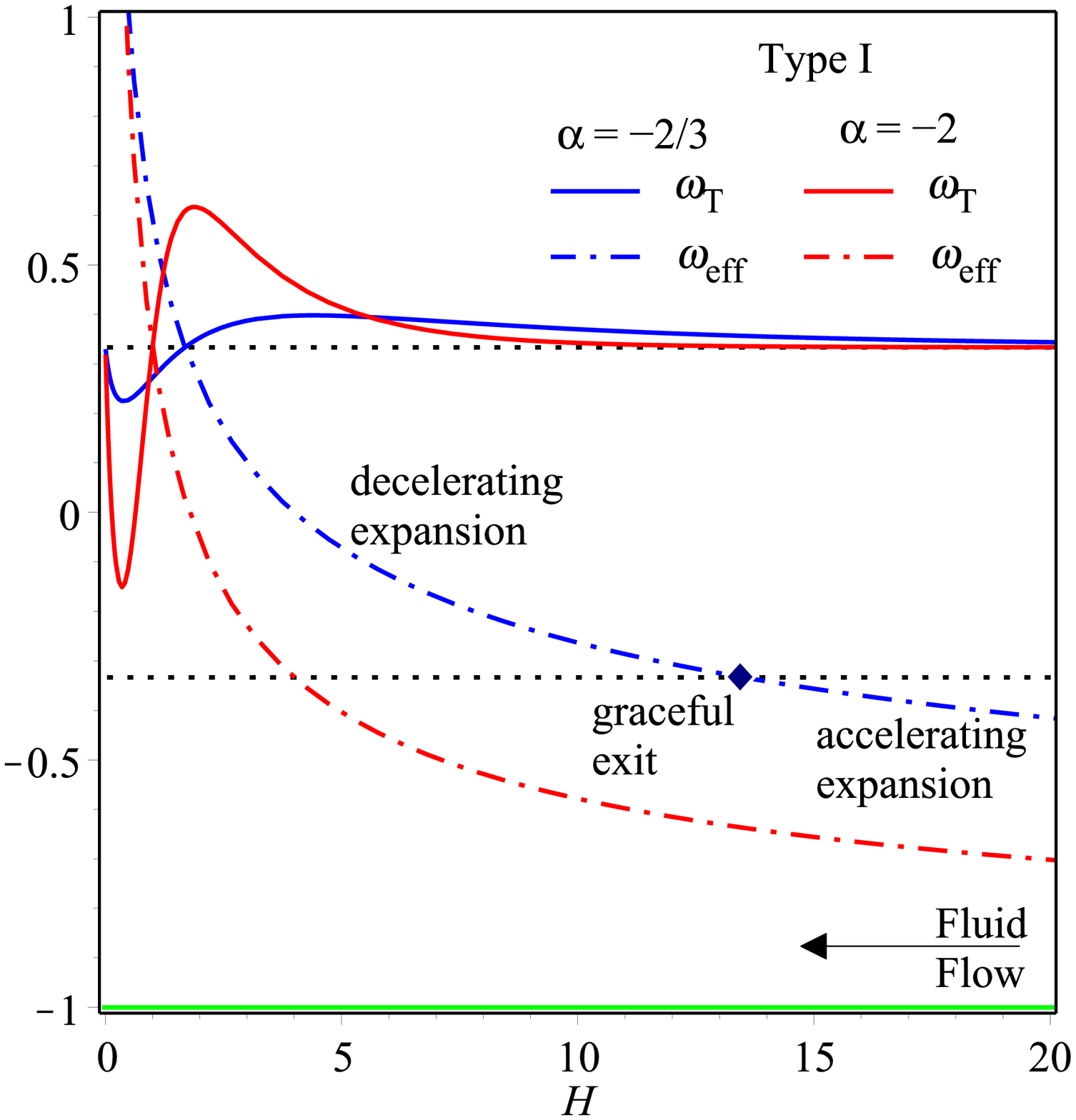}}
\caption[figtopcap]{The evolution of the effective and the torsion equation of state parameters, Eqs. (\ref{eff_EoS}) and (\ref{Tor_EoS}), in the vicinity of a finite-time singularity of Type I.}
\label{Fig:EoS-TypeI}
\end{figure}
\subsubsection{$f_{0}>0$ ($\beta<0$)}\label{Sec:4.1.1}
As we have discussed this case using the phase portrait, Fig. \ref{Fig:phspTypeI}\subref{fig:phsptypeIa}, the universe begins with a finite-time singularity of Type I in a contracting phase in phantom regime, and evolves towards a future fixed point $H\to 0$ in an infinite time. Using Eqs. (\ref{eff_EoS}) and (\ref{Tor_EoS}), we have, at the limit $H\to - \infty$, the universe effectively follows the torsion fluid as $\omega_\textmd{eff}=\omega_{T}\to -1$. At the limit $H\to 0$, the effectively equation of state diverges as $\omega_\textmd{eff}=Q_{0}\to -\infty$ but the universe needs an infinite time to approach that fate. Therefore, the universe will not feel this future singularity. On the contrary, the torsion fluid evolves towards $\omega_{T}\to -\infty$ similar to the effective fluid, but in a finite time. As seen in Fig. \ref{Fig:EoS-TypeI}\subref{fig:EoS-typeIa}, the irregular behavior of torsion fluid is just before the Minkowskian fixed point, whereas the torsion equation of state parameter changes its sign from $-\infty$ to $+\infty$. This behavior usually result in the sudden singularities. On another word, although the universe effectivly will not feel the singularity at the fixed point, the torsion fluid feels it as a finite-time singularity of Type II.  However, the torsion fluid crosses the phantom divide line through a singularity, then it evolving towards the matter equation of state, i.e. $\omega_{T}\to \omega_{m}$, as a final fate.

We note that the limits of Eqs. (\ref{eff_EoS}) and (\ref{Tor_EoS}) are both regular at $H\to +\infty$, where $\omega_\textmd{eff}=\omega_{T}\to -1$. Guided by the phase portraits of Fig. \ref{Fig:phspTypeI}, we should ignore these limits. However, we recorded them in Table \ref{Table:EoS} in Sec. \ref{Sec:5} just for completeness.
\subsubsection{$f_{0}<0$ ($\beta>0$)}\label{Sec:4.1.2}
In these models, the universe begins with a finite-time singularity of Type I (big bang) similar to the standard cosmology. However, it experienced an early accelerated expansion phase, then it enters a later decelerating phase. For more details, one may recall Sec. \ref{Sec:2.1.2}.\\

\textit{Case 1.}  In the subclass $-2<\alpha<-1$, the universe begins with a big bang singularity where the asymptotic behavior of the effective equation of state (\ref{eff_EoS}) can be obtained as $\omega_\textmd{eff}\to -1$ as $H\to \infty$. As seen in Fig. \ref{Fig:EoS-TypeI}\subref{fig:EoS-typeIb}, the universe effectively crosses $\omega_\textmd{eff}=-1/3$ ending the early inflationary phase, and enters a late deceleration phase as $\omega_\textmd{eff}>-1/3$. We note that the plots of the Fig. \ref{Fig:EoS-TypeI}\subref{fig:EoS-typeIb} are just to visualize the qualitative behavior of the model, for physical models, however, one may consult Fig. \ref{Fig:alpha-beta}\subref{fig:trans-typeI} to use the correct values of the parameters $\alpha$ and $\beta$ in order to end the inflation period at suitable energy scale $10^7<H<10^9$ GeV. Then, the effective equation of state diverges, i.e. $\omega_\textmd{eff}=Q_{0}\to +\infty$ as $H$ drops to zero, that is a fixed point and effective fluid takes an infinite time to reach that point. On the other hand, using Eq. (\ref{Tor_EoS}), we find that the torsion fluid matches the matter fluid during the cosmic time, $\omega_{T} \sim \omega_{m}$, with an oscillationary behavior at the limit $H\to 0$.\\

\textit{Case 2.}  In the subclass $\alpha\leq -2$, the evolution is very similar to case 1 above as seen in Fig. \ref{Fig:EoS-TypeI}\subref{fig:EoS-typeIb}. But we find that the torsion equation of state has an asymptotic behavior $\omega_{T}\to -1$ at the limit $H\to -\infty$ instead of $\omega_{m}$. This has been shown clearly in Table \ref{Table:EoS} in Sec. \ref{Sec:5}. However, the negative $H$ region should be excluded as guided by the phase portraits in Fig. \ref{Fig:phspTypeI}\subref{fig:phsptypeIb}. So we note that the two cases of $\beta>0$, in general, are identical.
\subsection{Type II singularity}\label{Sec:4.2}
\begin{figure}
\centering
\subfigure[~$\alpha=\frac{1}{n}$, $n$ = odd ($\beta<0$)]{\label{fig:EoS-typeIIa}\includegraphics[scale=.28]{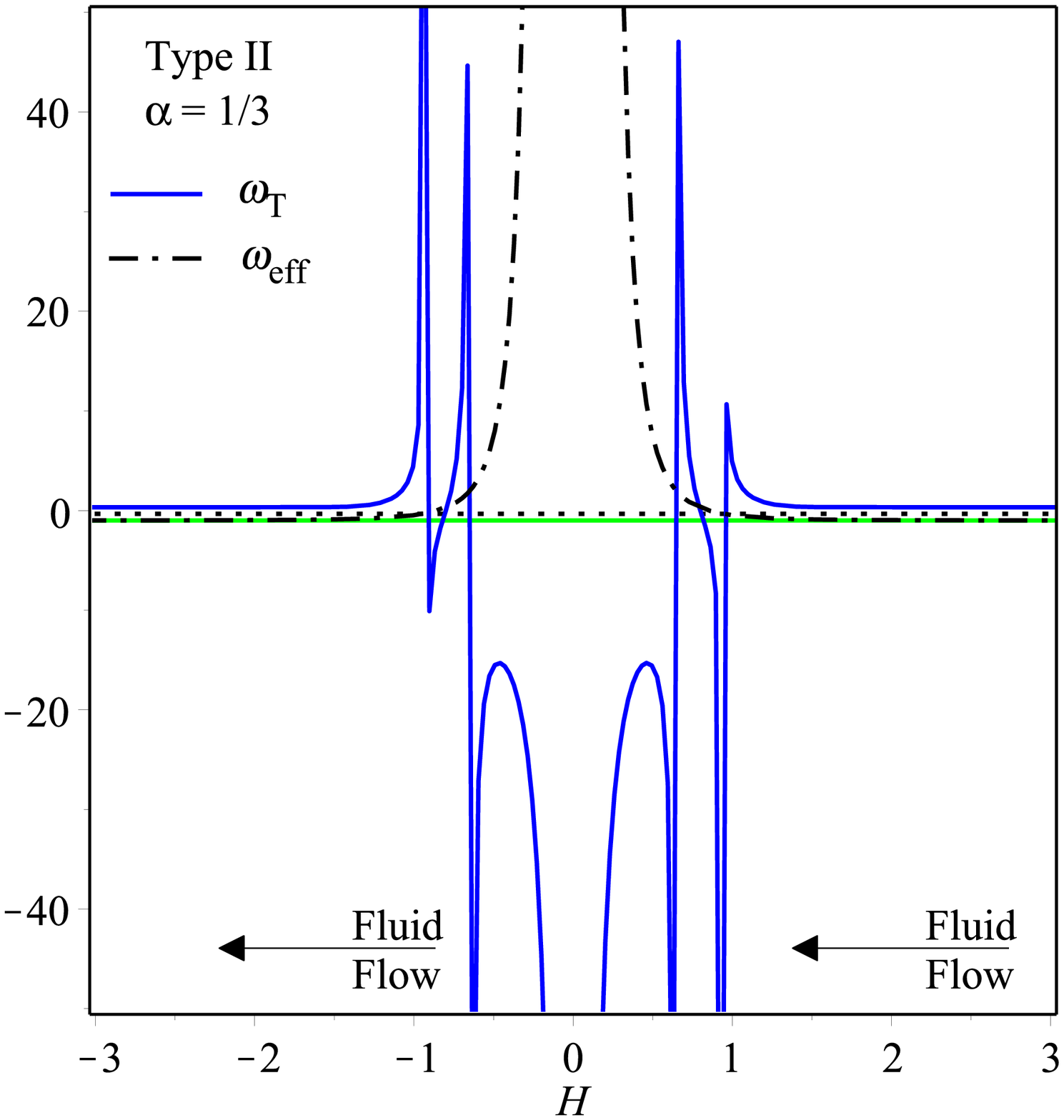}}
\subfigure[~$\alpha=\frac{1}{n}$, $n$ = even ($\beta<0$ or $\beta>0$)]{\label{fig:EoS-typeIIb}\includegraphics[scale=.29]{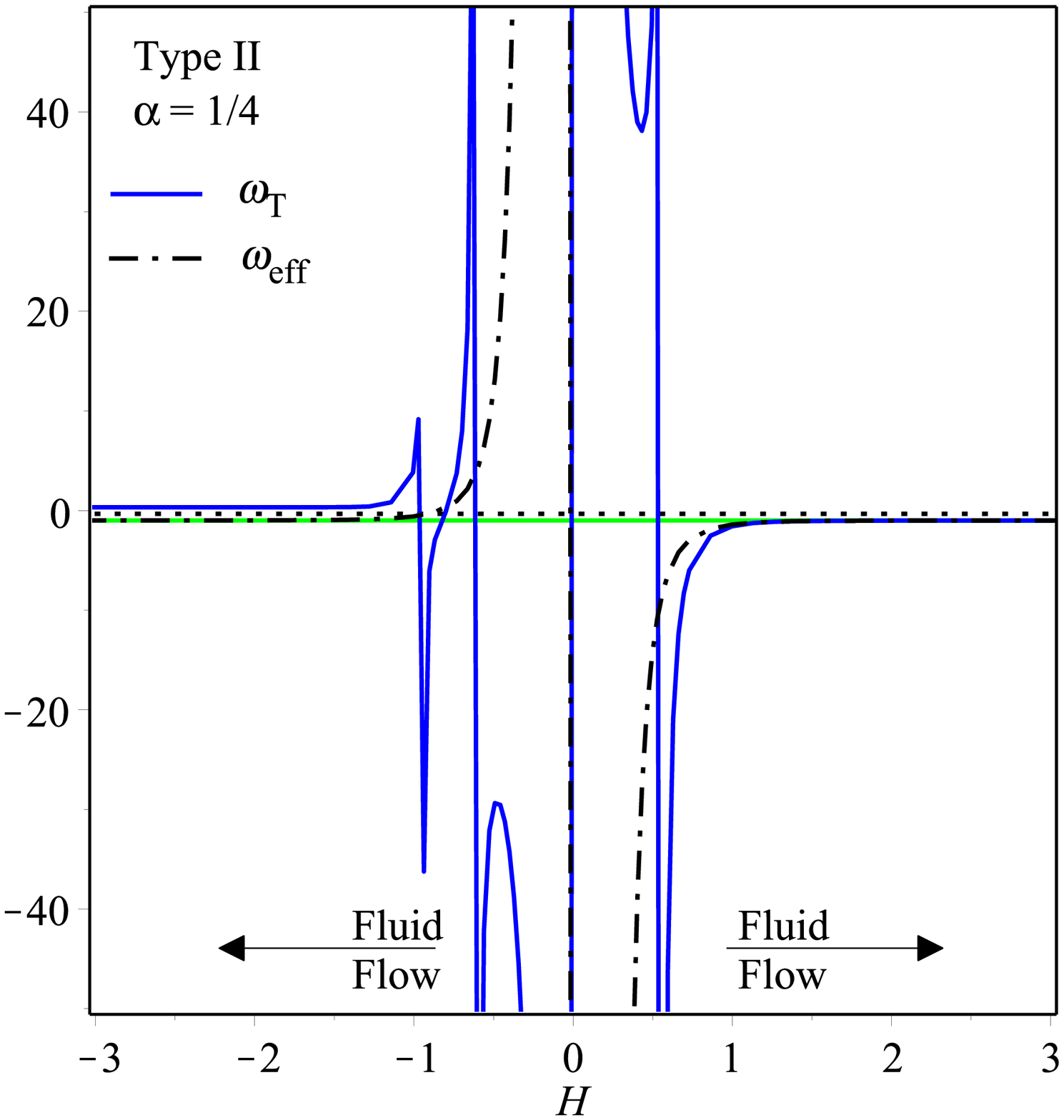}}
\subfigure[~$\alpha=\frac{1}{n}$, $n$ = odd ($\beta>0$)]{\label{fig:EoS-typeIIc}\includegraphics[scale=.28]{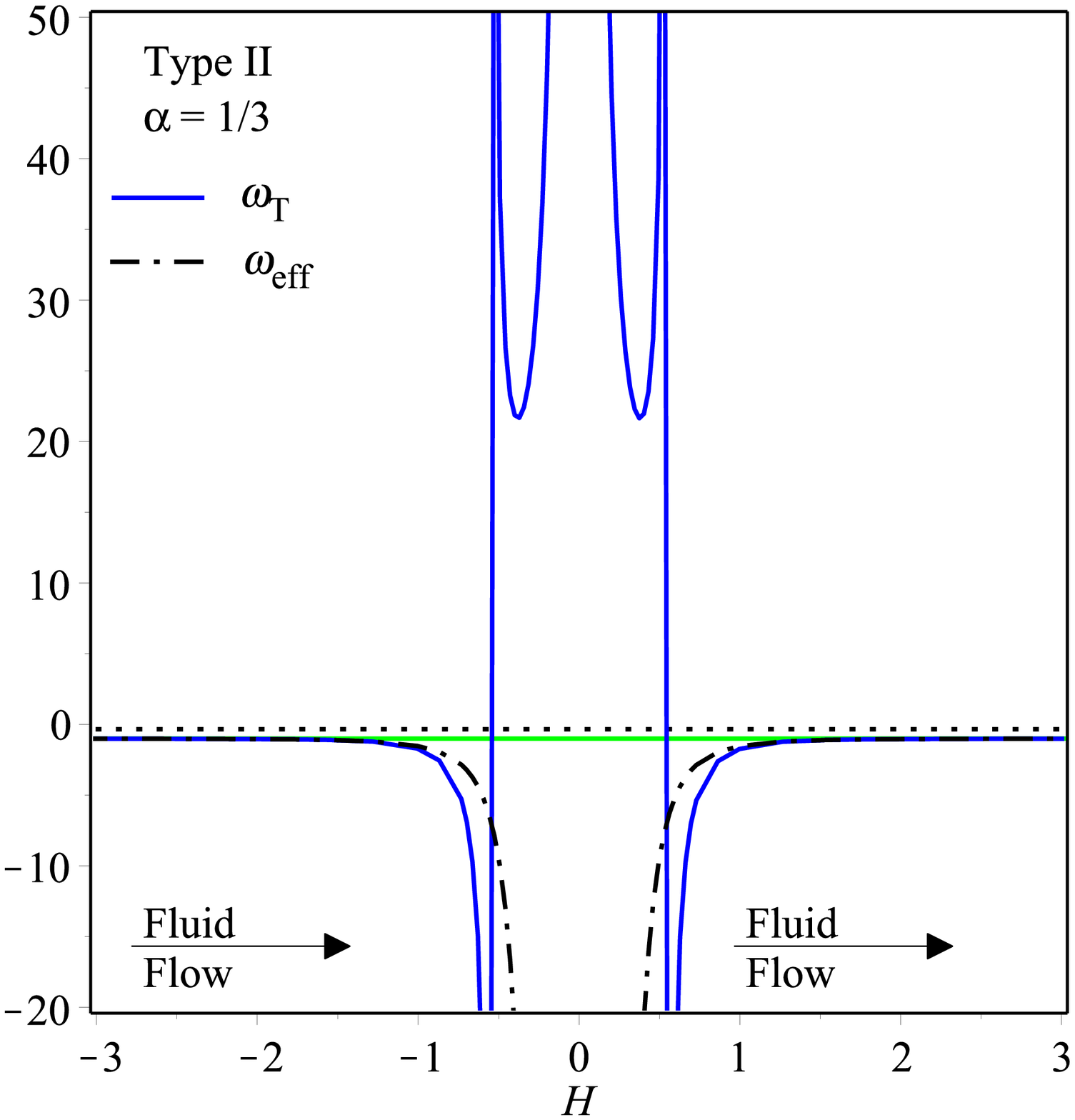}}
\caption[figtopcap]{The evolution of the effective and the torsion equation of state parameters, Eqs. (\ref{eff_EoS}) and (\ref{Tor_EoS}), in the vicinity of a finite-time singularity of Type II.}
\label{Fig:EoS-TypeII}
\end{figure}

Using Eqs. (\ref{eff_EoS}) and (\ref{Tor_EoS}), we plot their evolutions as given in Fig. \ref{Fig:EoS-TypeII}. As in Sec. \ref{Sec:2.2}, we discuss two main categories of the models which have singularity of Type II ($0<\alpha<1$), that are $\beta<0$ and $\beta>0$. Recalling the phase portraits of Fig. \ref{Fig:phspTypeII}, we note that the singularity is at $H_s=0$, where $\dot{H}$ diverges. This leads the effective pressure to diverge as well. Consequently, it will be common for all the following categories to have a divergent effective equation of state at that point, i.e. $\omega_\textmd{eff}\to \pm \infty$ as $H\to 0$.
\subsubsection{$f_0 <0$ ($\beta<0$)}\label{Sec:4.2.1}
\textit{Case 1.} In the subclass $\alpha=\frac{1}{n}$ (where ($n>1$ is an odd positive integer), we plot the evolution of both the effective equation of state and the torsion one as appear in Fig. \ref{Fig:EoS-TypeII}\subref{fig:EoS-typeIIa}. The evolution goes in the decreasing Hubble. Since the singularity is not of a crashing type and the geodesics are complete, a transition from expansion to contraction is conceivable. These can be called a big brake model, usually they suffer from a paradox of singularity crossing, where tachyon cosmological models \cite{Keresztes:2009vc} or anti-Chaplygin gas \cite{Keresztes:2012zn} play an essential role to make the singularity crossing physically possible. We note, for this model, that the torsion fluid asymptotically matches the matter component, i.e. $\omega_{T}\to \omega_{m}$. However, its equation of state feels the singularity earlier and becomes irregular. In the vicinity of the singularity $H\to 0^{\pm}$, from Eq. (\ref{Tor_EoS}), we have
\begin{equation}\label{Q1}
\lim_{H\to 0^{\pm}}\omega_T= Q_{1}\to -\infty;~\textmd{since}~\beta<0~\textmd{and}~0<\alpha<1,
\end{equation}
where $Q_{1}\equiv Q_{1}(H)=\frac{2}{\rho_{0}\kappa^2}\alpha H \left(\frac{\beta}{H}\right)^{1/\alpha}$. However, in the contraction phase the torsion fluid evolves towards the matter component.\\

\textit{Case 2.} In the subclass $\alpha=\frac{1}{n}$ (where ($n>1$ is an even positive integer), we plot the evolution of both the effective equation of state and the torsion one as appear in Fig. \ref{Fig:EoS-TypeII}\subref{fig:EoS-typeIIb}. The sudden singularity at $H=0$ acts as a repeller, whereas the universe evolves in the increasing of the Hubble parameter in the $H>0$ patch and it evolves in the decreasing Hubble in the $H<0$ patch. In this subclass, the effective fluid evolves in the vicinity of the singularity as
$$\lim_{H\to 0^{\pm}}\omega_\textmd{eff}=Q_{0}\to \mp \infty,$$
while it asymptotically evolves towards the cosmological constant, i.e. $\omega_\textmd{eff}\to -1$ as $H\to \pm \infty$.
At the right patch, the torsion evolves to cosmological constant, i.e. $\omega_T\to -1$ as $H\to +\infty$. At the left patch, it evolves towards the matter component, i.e. $\omega_T\to \omega_m$ as $H\to -\infty$. However, in the vicinity of the singularity $H=0$, the torsion equation of state alters its sign opposite to the effective fluid as
$$\lim_{H\to 0^{\pm}}\omega_{T}=Q_{1}\to \pm \infty.$$

\textit{Case 3.} In the subclass $\alpha=\frac{a}{b}\neq \frac{1}{n}$, where ($a$ and $b$ are positive integers such that $a<b$), we find out two patterns: The first is when $0<\alpha<1/2$, it follows case 2 and can be visualized on Fig. \ref{Fig:EoS-TypeII}\subref{fig:EoS-typeIIa}. The second is when $1/2<\alpha<1$, it follows case 1 and can be visualized on Fig. \ref{Fig:EoS-TypeII}\subref{fig:EoS-typeIIb}. For both cases only the $H<0$ patch is allowed for this subclass.
\subsubsection{$f_0 >0$ ($\beta>0$)}\label{Sec:4.2.2}
\textit{Case 1.} In the subclass $\alpha=\frac{1}{n}$ (where ($n>1$ is an odd positive integer), we plot the evolution of both the effective equation of state and the torsion one as appear in Fig. \ref{Fig:EoS-TypeII}\subref{fig:EoS-typeIIc}. This subclass can be used to describe a bouncing cosmology model, since crossing the singularity at $H=0$ from contraction ($H<0$) to expansion ($H>0$) is a valid scenario. In general, bouncing models suffer from two main problems: The ghost instability and the anisotropy grows. The first arises when the null energy condition has been broken. In the $f(T)$ gravity, it has been shown that the null energy condition is violated effectively only, while the matter component is free from forming ghost degrees of freedom \cite{Bamba2016}. The second problem arises during the contraction phase before bounce, since the anisotropies grow faster than the background so that the universe ends up to a complete anisotropic universe and bouncing to expansion will not occur.

As shown in Fig. \ref{Fig:EoS-TypeII}\subref{fig:EoS-typeIIc}, the effective fluid acts asymptotically $H\to \pm \infty$ as a cosmological constant $\omega_\textmd{eff}\to -1$, then it runs deeply in phantom regime near the singular bounce as
$$\lim_{H\to \pm}\omega_\textmd{eff}=Q_{0}\to -\infty.$$
However, the torsion fluid in the vicinity of the bouncing (singular) point has a large equation of state,
$$\lim_{H\to 0^{\pm}}\omega_{T}=Q_{1}\to +\infty,$$
which allows the torsion gravity background to dominate over the anisotropy during the contraction avoiding the remain main problem of bouncing models. Therefore, the model at hand can be considered as a healthy bouncing scenario, where the torsion gravity plays the main role to avoid the usual problems of the bounce cosmological models.\\

\textit{Case 2.} In the subclass $\alpha=\frac{1}{n}$ (where ($n>1$ is an even positive integer), the evolution is identical to that has been obtained in case 2 of Sec. \ref{Sec:4.2.1}. So the evolution can be realized from Fig. \ref{Fig:EoS-TypeII}\subref{fig:EoS-typeIIb}.\\

\textit{Case 3.} In the subclass $\alpha=\frac{a}{b}\neq \frac{1}{n}$, where ($a$ and $b$ are positive integers such that $a<b$), we find two patterns: The first is when $0<\alpha<1/2$, it follows case 2 and can be visualized on Fig. \ref{Fig:EoS-TypeII}\subref{fig:EoS-typeIIc}. The second is when $1/2<\alpha<1$, it follows case 1 and can be visualized on Fig. \ref{Fig:EoS-TypeII}\subref{fig:EoS-typeIIb}. For both only the $H>0$ patch, where the $H<0$ patch is not allowed for this subclass.\\

We note that the asymptotic behavior of both the effective and the torsion equations of state and in the vicinity of the Type II finite-time singularity is summarized in Table \ref{Table:EoS} in Sec. \ref{Sec:5}.
\subsection{Type III singularity}\label{Sec:4.3}
Using Eqs. (\ref{eff_EoS}) and (\ref{Tor_EoS}), we plot their evolutions as given in Fig. \ref{Fig:EoS-TypeIII}. As in Sec. \ref{Sec:2.2}, we discuss two main categories of the models which have singularity of Type III ($-1<\alpha<0$), that are $\beta<0$ and $\beta>0$:
\subsubsection{$f_0<0$ ($\beta<0$)}\label{Sec:4.3.1}
\begin{figure}
\centering
\subfigure[~$\alpha=-\frac{1}{n}$, $n$ = odd ($\beta<0$)]{\label{fig:EoS-typeIIIa}
\includegraphics[scale=.28]{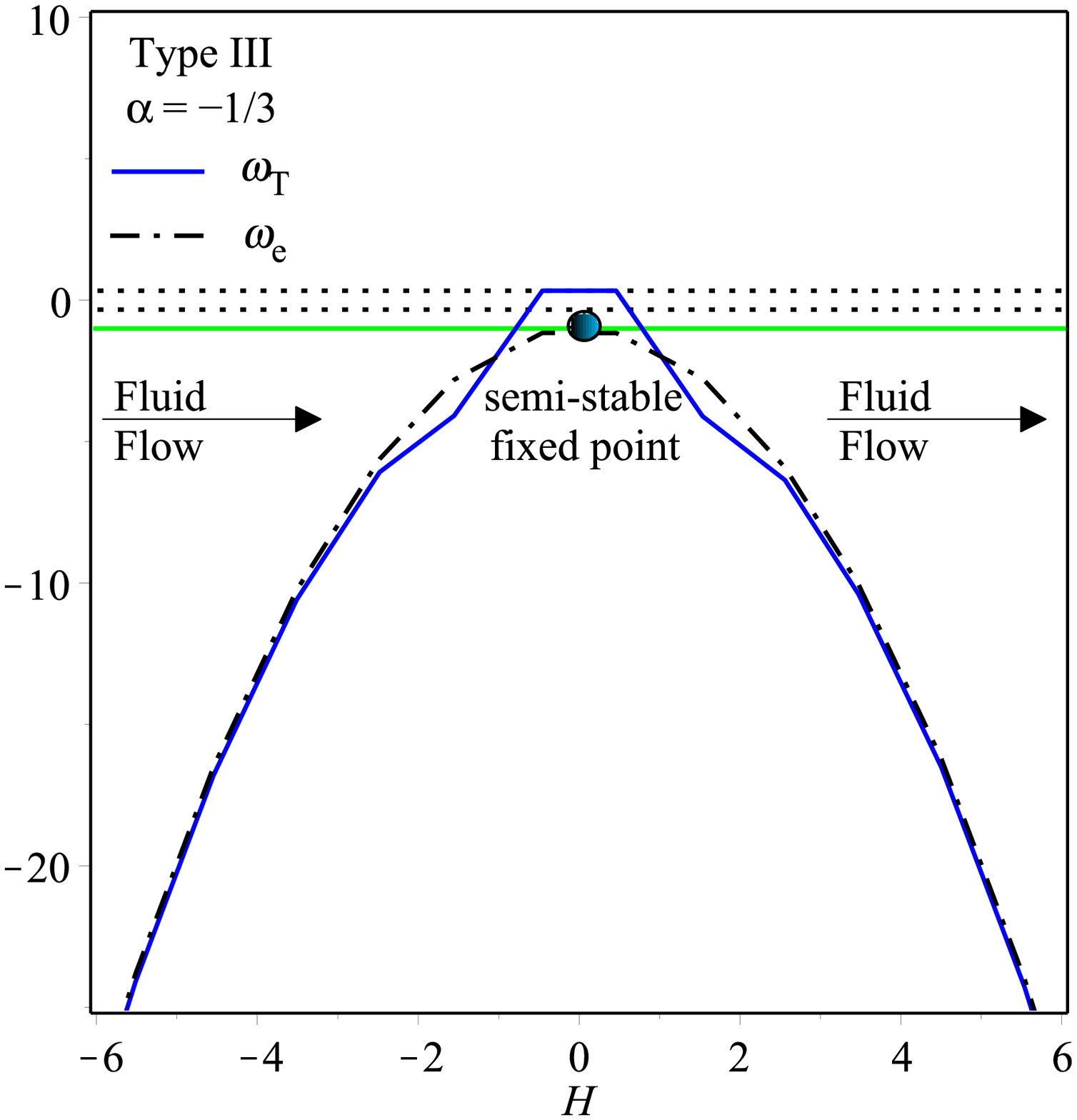}}
\subfigure[~$\alpha=-\frac{1}{n}$, $n$ = even ($\beta<0$ or $\beta>0$)]{\label{fig:EoS-typeIIIb}
\includegraphics[scale=.28]{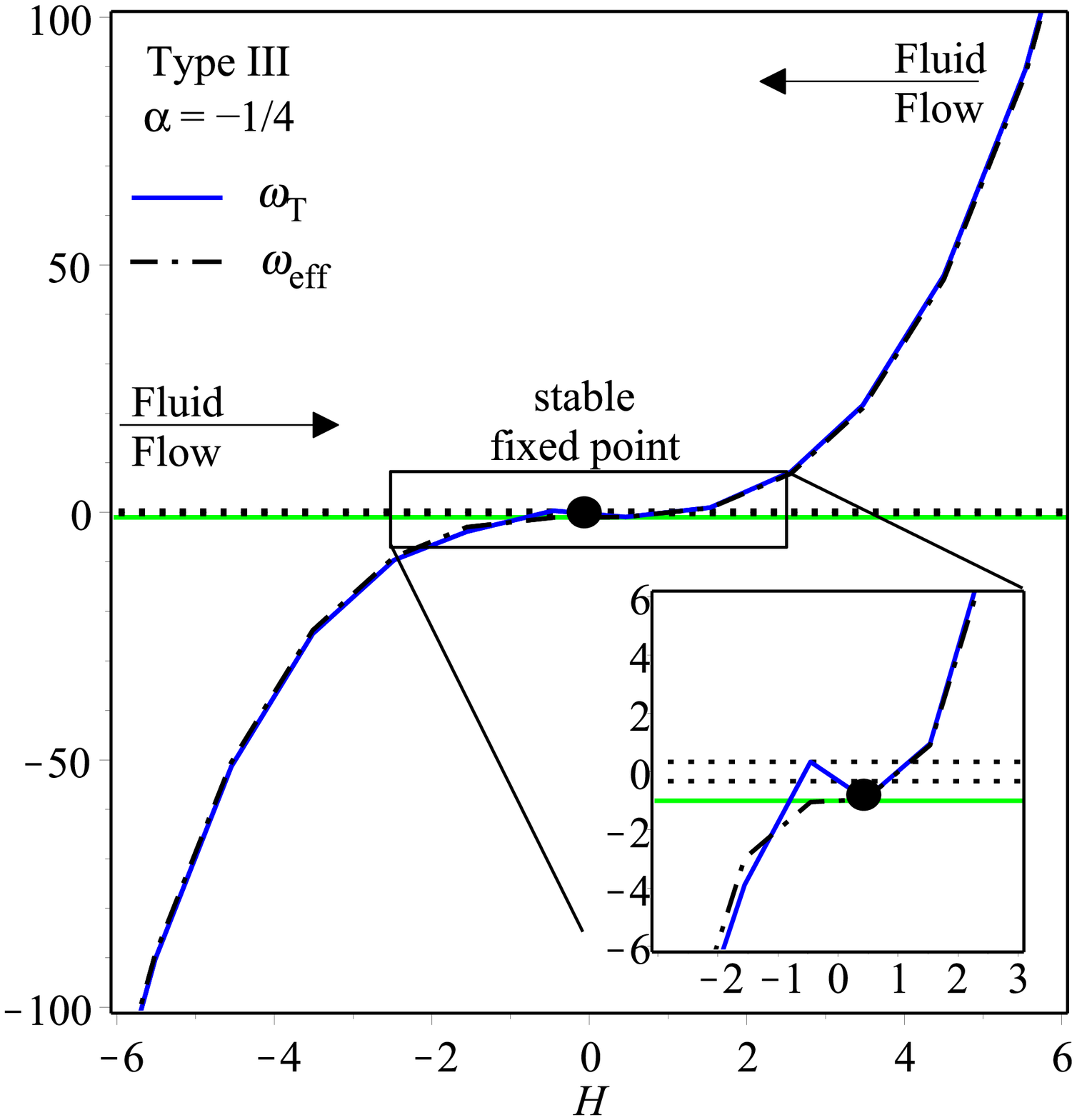}}
\subfigure[~$-1<\alpha=-\frac{a}{b}\neq -\frac{1}{n}<-1/2$ ($\beta<0$)]{\label{fig:EoS-typeIIIc}
\includegraphics[scale=.28]{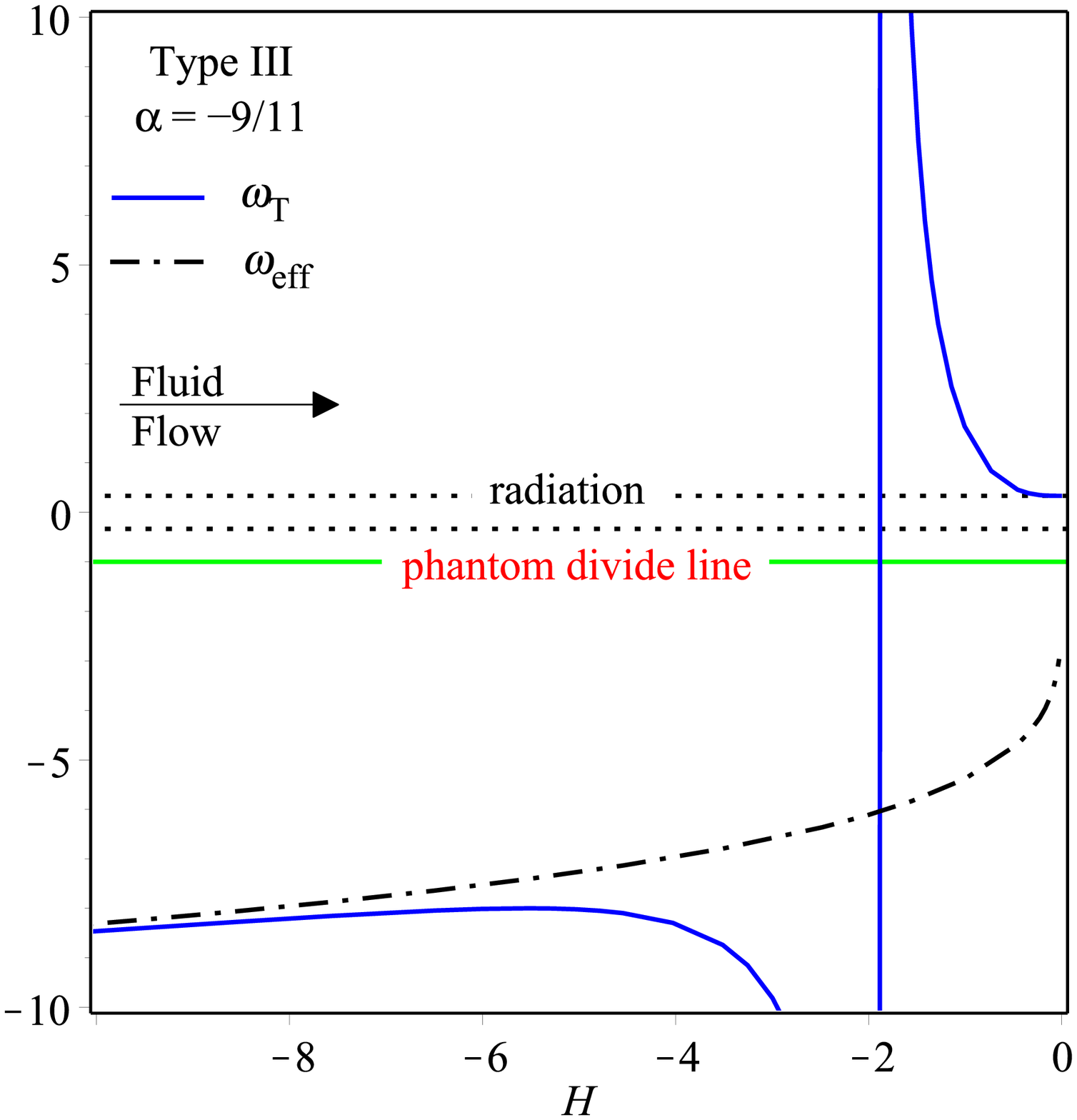}}\\
\subfigure[~$\alpha=-\frac{1}{n}$, $n$ = odd ($\beta>0$)]{\label{fig:EoS-typeIIId}
\includegraphics[scale=.28]{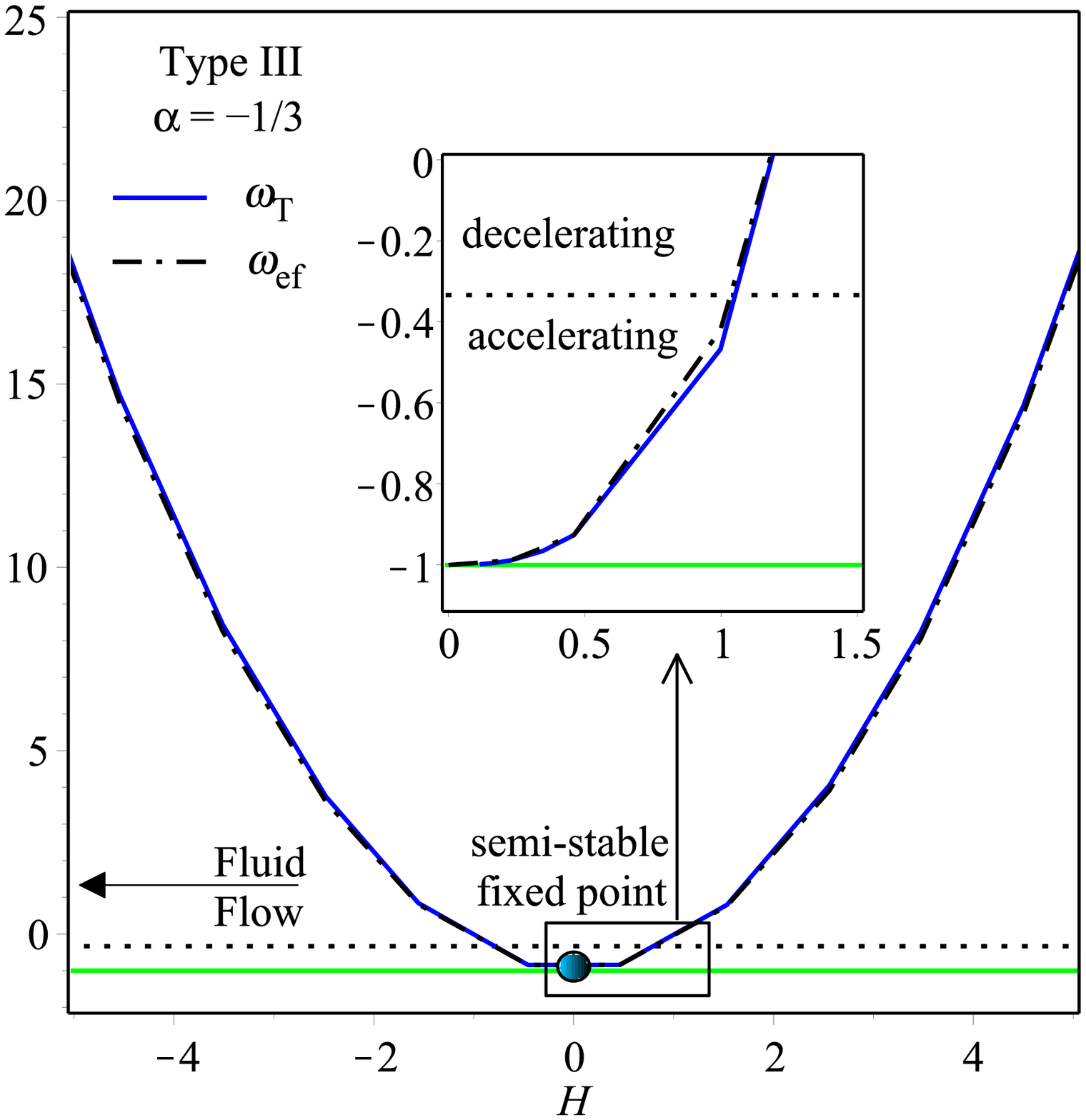}}
\subfigure[~$-1<\alpha=-\frac{a}{b}\neq -\frac{1}{n}<-1/2$ ($\beta>0$)]{\label{fig:EoS-typeIIIe}
\includegraphics[scale=.28]{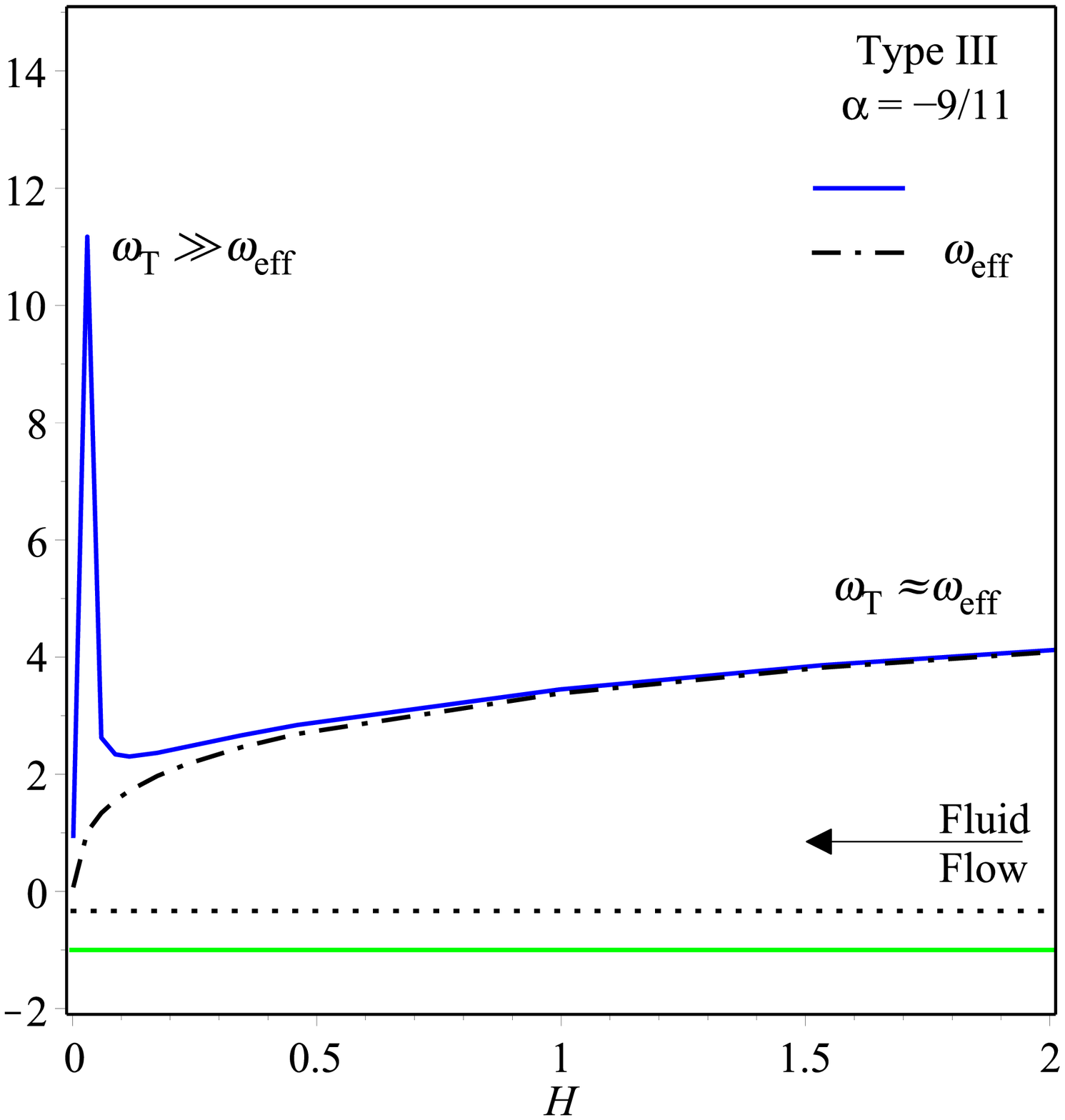}}
\caption[figtopcap]{The evolution of the effective and the torsion equation of state parameters, Eqs. (\ref{eff_EoS}) and (\ref{Tor_EoS}), in the vicinity of a finite-time singularity of Type III.}
\label{Fig:EoS-TypeIII}
\end{figure}

\textit{Case 1.} In the subclass $\alpha=-\frac{1}{n}$ (where ($n>1$ is an odd positive integer), we plot the evolution of both the effective and the torsion equations of state as appear in Fig. \ref{Fig:EoS-TypeIII}\subref{fig:EoS-typeIIIa}. From Eqs. (\ref{eff_EoS}) and (\ref{Tor_EoS}), we find that both equation of state parameters, asymptotically, evolve as $\omega_\textmd{eff}=\omega_{T}= Q_{0}\to -\infty$ as $H\to \pm \infty$. However, as $H\to 0$, the effective fluid evolves towards a cosmological constant, i.e $\omega_\textmd{eff}\to -1$, while the torsion fluid evolves towards the matter component, i.e. $\omega_{T}\to \omega_{m}$. On another word, the universe evolves effectively in a phantom regime towards the cosmological constant, while the torsion fluid crosses the phantom divide line towards the matter component at that point. We note that the point $H=0$ is a fixed point so the fluids reaches the mentioned fate in an infinite time, while asymptotically there are finite-time singularities of Type III, and therefore, they reach their fate in a finite time. Also, the evolution in both patches, $H<0$ and $H>0$, occur in the phantom regime as $\omega_\textmd{eff}<-1$. This is in agreement with the corresponding phase portraits of Fig. \ref{Fig:phspTypeIII}\subref{fig:phsptypeIIIa}.\\

\textit{Case 2.} In the subclass $\alpha=-\frac{1}{n}$ (where ($n>1$ is an even positive integer),  we plot the evolution of both the effective and the torsion equations of state as appear in Fig. \ref{Fig:EoS-TypeIII}\subref{fig:EoS-typeIIIb}. From Eqs. (\ref{eff_EoS}) and (\ref{Tor_EoS}), noting that $n$ is even and $\beta<0$ in the case at hand, the sign of the asymptotic behavior will be sensitive only to the sign of the Hubble parameter. Thus, we write
$$\lim_{H\to \pm \infty} \omega_\textmd{eff}=\lim_{H\to \pm \infty} \omega_{T}= Q_{0}\to \pm \infty.$$
However, at the limit $H\to 0$ both fluids evolve towards the cosmological constant, i.e. $\omega_\textmd{eff}=\omega_{T}\to -1$. As clear from Fig. \ref{Fig:EoS-TypeIII}\subref{fig:EoS-typeIIIb}, in the $H<0$ patch the universe evolves effectively in the phantom regime towards a cosmological constant. However, in the $H>0$ patch the universe evolves effectively in a non phantom regime towards the cosmological constant at the Minkowski fixed point universe, so it realizes a late transition to accelerating expansion universe as it crosses from $\omega_\textmd{eff}>-1/3$ to $-1<\omega_\textmd{eff}<-1/3$. Although, we can choose suitable values of $\alpha$ and $\beta$, as shown in Sec. \ref{Sec:2.5}, to realize the late transition at redshift $z_{de}\sim 0.7$ as required by the $\Lambda$CDM model, the universe evolves towards the Minkowski universe not de Sitter. This is not usual when dark energy is interpreted as a cosmological constant.\\

\textit{Case 3.} In the subclass $\alpha=-\frac{a}{b}\neq -\frac{1}{n}$, where ($a$ and $b$ are positive integers such that $a<b$), we find two patterns: The first is when $-1/2<\alpha<0$, it follows case 2 and can be visualized on Fig. \ref{Fig:EoS-TypeIII}\subref{fig:EoS-typeIIIa}, but only the $H<0$ patch, where the $H>0$ patch is not allowed for this subclass. The second is when $-1<\alpha<-1/2$, the evolution at the limits $H\to 0,~\pm \infty$ follows case 1 as given in Table \ref{Table:EoS} in Sec. \ref{Sec:5}. However, we plot the instantaneous evolution of both the effective equation of state and the torsion one as appear in Fig. \ref{Fig:EoS-TypeIII}\subref{fig:EoS-typeIIIc}, whereas the $H<0$ patch is the only valid scenario.
\subsubsection{$f_0>0$ ($\beta>0$)}\label{Sec:4.3.2}
\textit{Case 1.} In the subclass $\alpha=-\frac{1}{n}$ (where ($n>1$ is an odd positive integer), we plot the evolution of both the effective and the torsion equations of state as appear in Fig. \ref{Fig:EoS-TypeIII}\subref{fig:EoS-typeIIId}. From Eqs. (\ref{eff_EoS}) and (\ref{Tor_EoS}), we find that both equation of state parameters, asymptotically, evolve as $\omega_\textmd{eff}=\omega_{T}= Q_{0}\to +\infty$ as $H\to \pm \infty$. This is opposite to the similar case of Sec. \ref{Sec:4.3.1} as we examine $\beta>0$ in this section. However, as $H\to 0$, both parameters evolve towards the cosmological constant, i.e $\omega_\textmd{eff}=\omega_{T}\to -1$. Thus, the universe evolves effectively in a non phantom regime towards the cosmological constant, where a late transition from acceleration to deceleration is valid. This is in agreement with the corresponding phase portraits of Fig. \ref{Fig:phspTypeIII}\subref{fig:phsptypeIIIa}.\\

\textit{Case 2.} In the subclass $\alpha=-\frac{1}{n}$ (where ($n>1$ is an even positive integer),  the evolution is identical to that has been obtained in case 2 of Sec. \ref{Sec:4.3.1}. So the evolution can be realized from Fig. \ref{Fig:EoS-TypeIII}\subref{fig:EoS-typeIIIb}. \\

\textit{Case 3.} In the subclass $\alpha=-\frac{a}{b}\neq -\frac{1}{n}$, where ($a$ and $b$ are positive integers such that $a<b$), we find two patterns: The first is when $-1/2<\alpha<0$, it follows case 2 and can be visualized on Fig. \ref{Fig:EoS-TypeIII}\subref{fig:EoS-typeIIId}, but only the $H>0$ patch, where the $H<0$ patch is not allowed for this subclass. The second is when $-1<\alpha<-1/2$, we plot the evolution of both the effective equation of state and the torsion one as appear in Fig. \ref{Fig:EoS-TypeIII}\subref{fig:EoS-typeIIIe}. From Eqs. (\ref{eff_EoS}) and (\ref{Tor_EoS}), we find that both equation of state parameters, asymptotically, evolve as
$$\lim_{H\to \pm \infty}\omega_\textmd{eff}=\lim_{H\to \pm \infty}\omega_{T}=Q_{0}\to \pm \infty,$$
whereas the $H<0$ patch is not allowed for this class. At the limit $H\to 0$, both parameters evolves towards the cosmological constant. However, the torsion fluid parameter could be irregular in the vicinity of that point.\\

We mention that all the above three cases where $\beta>0$, in addition to case 2 where $\beta<0$, can perform a late deceleration to acceleration transition. For possible suitable choices of the parameters $\alpha$ and $\beta$, one should recall Sec. \ref{Sec:2.5}, in particular Fig. \ref{Fig:alpha-beta}\subref{fig:trans-typeIII}.
\subsection{Type IV singularity}\label{Sec:4.4}
Using Eqs. (\ref{eff_EoS}) and (\ref{Tor_EoS}), we plot their evolutions as given in Fig. \ref{Fig:EoS-TypeIV}. As we have shown earlier in Sec. \ref{Sec:2.4}, there are two cases can be discussed for the models which have singularity of Type IV ($\alpha>1$):
\subsubsection{$f_{0}>0$ ($\beta<0$)}\label{Sec:4.4.1}
\begin{figure}
\centering
\subfigure[~$f_0>0$ ($\beta<0$)]{\label{fig:EoS-typeIVa}\includegraphics[scale=.28]{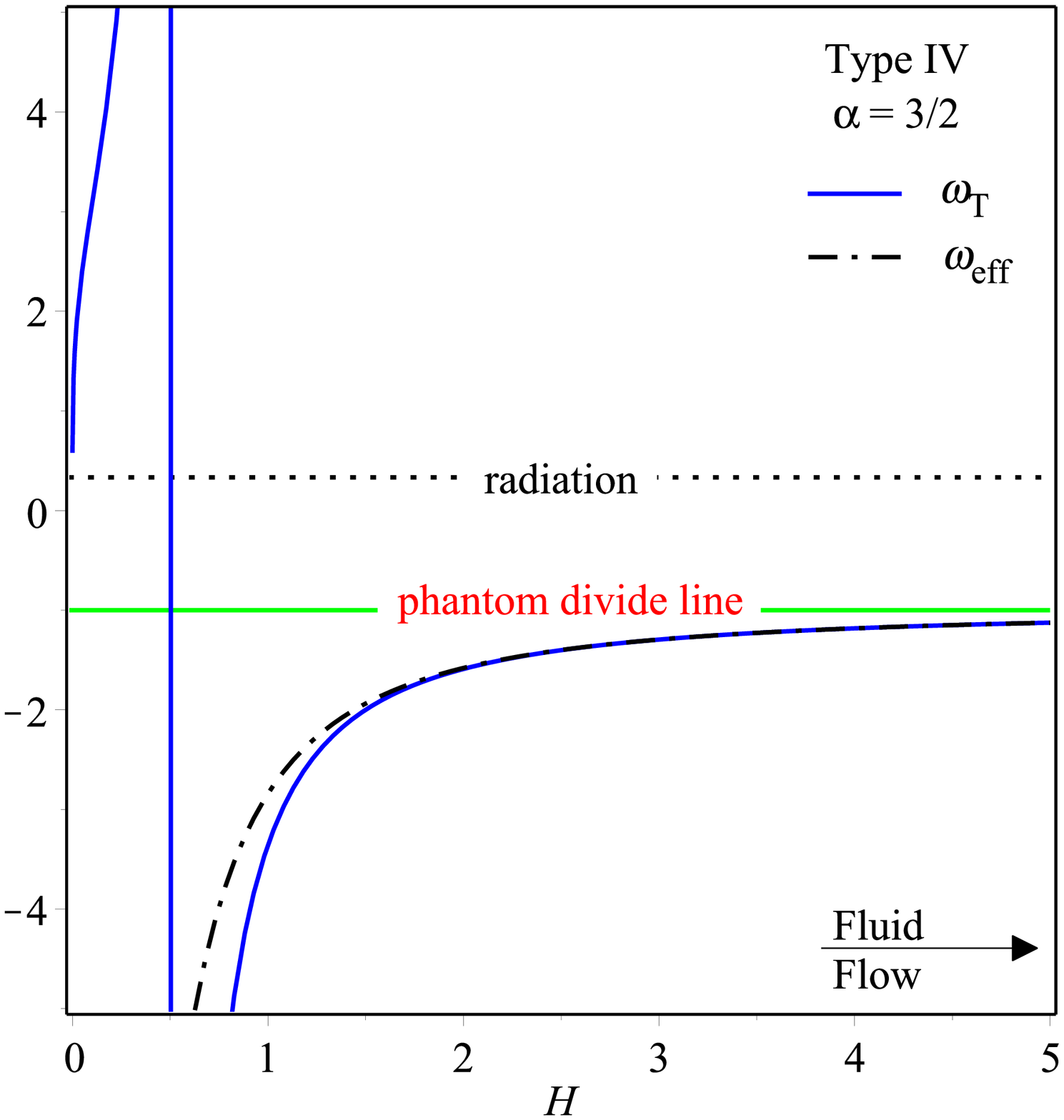}}\hspace{2mm}
\subfigure[~$f_0<0$ ($\beta>0$)]{\label{fig:EoS-typeIVb}\includegraphics[scale=.28]{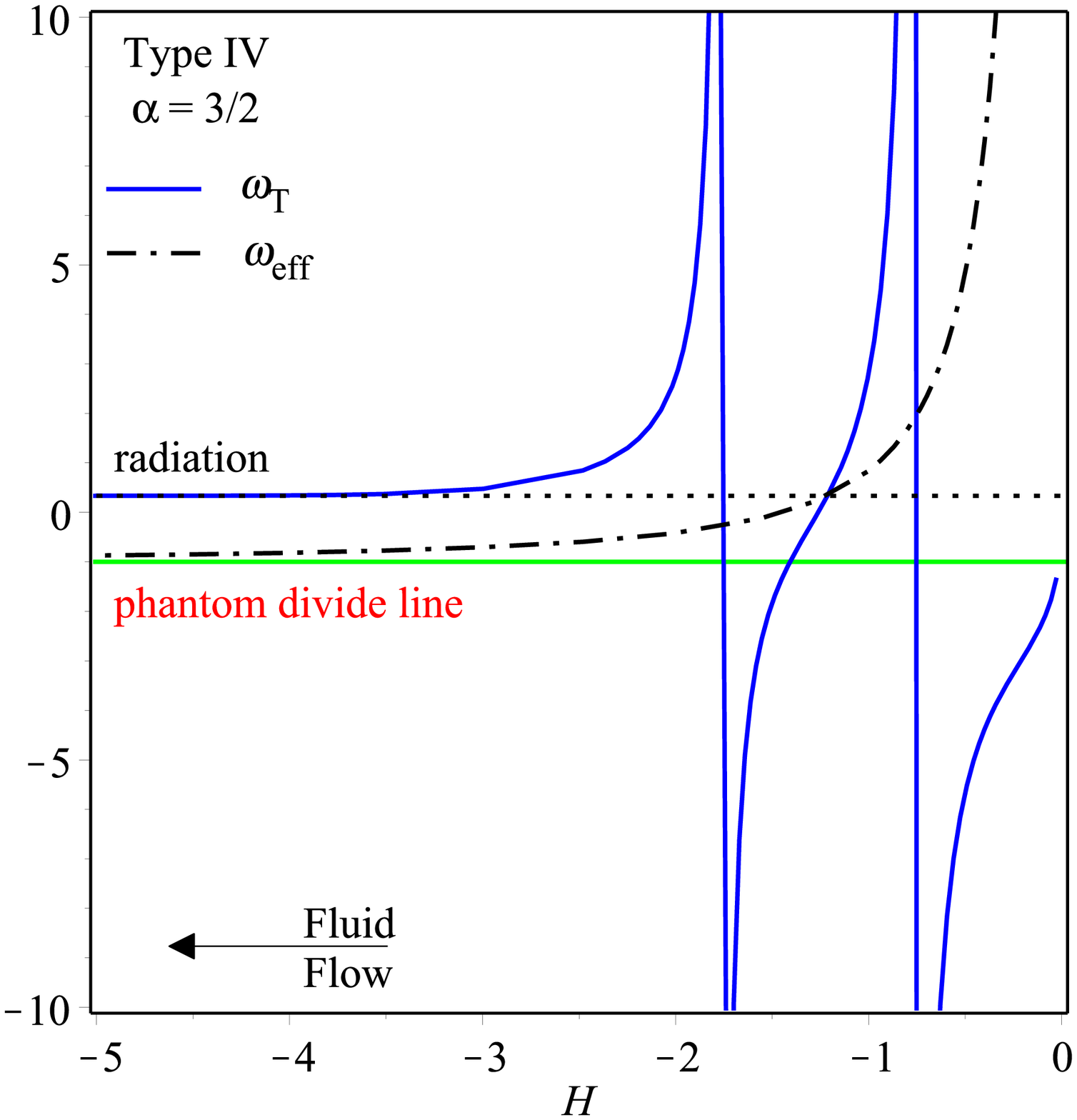}}
\caption[figtopcap]{The evolution of the effective and the torsion equation of state parameters, Eqs. (\ref{eff_EoS}) and (\ref{Tor_EoS}), in the vicinity of a finite-time singularity of Type IV.}
\label{Fig:EoS-TypeIV}
\end{figure}

As we have discussed this case using the phase portrait, Fig. \ref{Fig:phspTypeIV}, the universe begins with a finite-time singularity of Type IV at the Minkowski space which is a fixed point as well, therefore, the corresponding effective equation of state parameter has an infinite value at the Minkowiskian point. From Eq. (\ref{eff_EoS}), we have
$$\lim_{H\to 0^{+}}\omega_\textmd{eff}=Q_{0}=-\infty.$$
However, the torsion fluid at the same limit begins with an equation of state parameter equivalent to the matter component, i.e. $\omega_T\to \omega_{m}$ as $H\to 0$. Then the effective fluid evolves, in phantom regime, smoothly towards the cosmological constant, $\omega_\textmd{eff}\to -1$, in increasing $H$ direction. On the other hand, the torsion equation of state parameter matches the effective fluid asymptotically as $H\to +\infty$, it is irregular in the vicinity of the initial singularity. Since $\omega_T$ changes its sign near the singularity, it feels the singularity of Type IV as if it is a sudden singularity, see Fig. \ref{Fig:EoS-TypeIV}\subref{fig:EoS-typeIVa}. Recalling Eq. (\ref{sing-time}) and the related discussion, we find that the effective fluid will not be singular at $H\to \infty$, and consequently the torsion fluid as $\omega_\textmd{eff}\sim \omega_{T}$ at that limit.
\subsubsection{$f_{0}<0$ ($\beta>0$)}\label{Sec:4.4.2}
Alternatively, in these models, the universe begins effectively with $\omega_\textmd{eff}\to +\infty$ at a singular Minkowskian universe of Type IV. However, the torsion equation of state parameter initially begins with a value equivalent to the matter component, see Fig. \ref{Fig:EoS-TypeIV}\subref{fig:EoS-typeIVa}. Noting that only $H<0$ is the allowed patch in this scenario, so the universe is contracting with a deceleration after the Minkowskian universe as $\omega_\textmd{eff}>-1/3$, then it crosses the quintessence limit towards the cosmological constant asymptotically as $H\to -\infty$. On the other hand, the torsion fluid matches the matter component at that limit.

We note that the torsion fluid in the two subclasses $-2<\alpha<-1$ and $\alpha\leq -2$ having different limits at $H\to +\infty$. In the first subclass, the torsion fluid evolves asymptotically towards the matter component, while it evolves towards the cosmological constant in the second. However, the positive $H$ region should be excluded for $\beta>0$ as guided by the phase portraits in Fig. \ref{Fig:phspTypeIV}. So we note that the two cases of $\beta>0$, in general, are identical.\\

Finally, as clear from the discussions throughout the work at hand that investigating the asymptotical solutions and at the fixed points is essential to understand the one dimensional autonomous dynamical systems. We summarize all the limits of the effective and the torsion equation of state parameters at $H\to 0,~\pm \infty$ as given in Table \ref{Table:EoS} in Sec. \ref{Sec:5}.
\section{Concluding Remarks}\label{Sec:5}
For a modified gravity theory and by assuming the effective fluid has a linear equation of state, we have shown that the Friedmann equations can represent a one dimensional autonomous system, if the first derivative of Hubble would have been written as a function of the Hubble parameter only. This allows to interpret the Friedmann system as a vector field on a line introducing one of the basic techniques of dynamics. Consequently, we draw the ($H - \dot{H}$) phase space allowing a clear geometrical representation of the dynamical system by identifying its fixed points and the asymptotic behavior.\\

In this paper we investigated the phase portraits of the different types of the finite-time singularities. We adopted a scale factor containing two model parameters $\alpha$ and $\beta$, which can realize the four finite-time singularity types. The generic analysis of the phase portraits shown that:

(i) For the singularity of Type I ($\alpha<-1$), the model realizes an early accelerating expansion (inflation) with possible transition to a later decelerated expansion for some suitable values $\beta>0$. We plot the viable values of $\alpha$ and $\beta$ which allow this transition to occur at a cosmic energy scale $10^9$ to $10^7$ GeV when the inflation is assumed to be during $10^{-34}$ and $10^{-31}$ s. This provides a suitable conditions to have a graceful exit inflation model.

(ii) For the singularity of Type II ($0<\alpha<1$), when $\alpha$ is chosen such that $\alpha=\frac{1}{n}$ and $n$ is a positive integer, the universe evolves effectively in phantom regime. The universe realizes a contraction before the singularity, while it expand after it. This may gives a singular bounce with a finite-time singularity at the bouncing time. However, junction conditions could be used to overcome the singularity by welding the contraction an the expansion phases around the singular point. This point needs a further investigation.

(iii) For the singularity of Type III ($-1<\alpha<0$), the model realizes a late transition from decelerating to accelerating expansion.  We plot the viable values of $\alpha$ and $\beta$ which allow this transition to occur at redshift $z_{de}=0.72\pm 0.05$ ($0.84\pm 0.03$), i.e. $H_{de}=101.8$ ($117.82$) km/s/Mpc. Although, the model realizes the transition to the late accelerating expansion phase, it evolves towards a Minkowski fate not de Sitter as expected by the $\Lambda$CDM models. This is unusual when dark energy is assumed to be a cosmological constant.

(iv) For the singularity of Type IV ($\alpha>1$), remarkable the singular point is also a repeller fixed point, which provides an unstable de Sitter universe. In this case, we expect an unusual behavior by having a fixed point can be reached at a finite time. The phase portrait analysis show that the Type IV singularities can be used for crossing between phantom and non phantom cosmologies safely as the geodesics in this model are completed.\\
We also shown that the torsion based gravity is compatible with the phase portrait analysis. This is because the teleparallel torsion $T$ is a function of the Hubble parameter $H$ only and it contains no higher derivatives of $H$. Consequently, it allows the modified Friedmann system to be written as a one dimensional autonomous system. Among several versions of the generalized teleparallel gravity, we investigated the $f(T)$ gravity theories.\\

We reconstructed the $f(T)$ function which generates the proposed phase portraits. This allows to perform a complementary investigation of the singularity types through the torsion equation of state. In addition, we note that since some cases of the singularities are in fact a Minkowiskian space rather a de Sitter one, an investigation of the torsion role in the vicinity of different singularity types on the perturbation level will be needed, for perturbation analysis in $f(T)$ see \cite{Cai:2011tc,Chen:2010va,Karami,Bamba2016}. In fact the singular phases can be shifted from Minkowski to de Sitter as needed by modifying the phase portrait (\ref{autonomous}) by adding a constant $H_{ds}$. As we have shown, the dynamics of the one dimensional autonomous systems is dominated by the stability of the fixed points and the asymptotic behavior, we summarize below the obtained results of Sec. \ref{Sec:4} in Table \ref{Table:EoS}.
\begin{table}
\caption{The evolution of the effective/torsion equations of state (\ref{eff_EoS}) and (\ref{Tor_EoS}) near $H\to 0,~\pm \infty$ corresponding to different choices of $\alpha$ and $\beta$ parameters. We take $n>1$ ($m>1$) is a positive odd (even) integer, and $a<b$ where $a$ and $b$ are positive integers.}
\begin{center}\begin{tabular}{|c|c|c|c|c|c|c|c|c|c|}
\hline\hline
\multicolumn{4}{|c|}{parameters}&\multicolumn{3}{c|}{$\omega_\textmd{eff}$}          & \multicolumn{3}{c|}{$\omega_{T}$}\\
\hline
singularity &$f_{0}$ &$\beta$ & $\alpha$      & $H\to 0^{\pm}$ & $H\to \infty$ & $H\to -\infty$ & $H\to 0^{\pm}$ & $H\to \infty$ & $H\to -\infty$\\
\hline
{\multirow{3}{*}{\rotatebox[origin=c]{90}{Type I}}}&$>0$&$<0$&$<-1$&$Q_{0}\to -\infty$&$-1$&$-1$&$\omega_{m}$&$-1$&$-1$\\ \cline{2-10}
&\multirow{2}{*}{$<0$}&\multirow{2}{*}{$>0$}&$\in (-2,-1)$&$Q_{0}\to +\infty$&$-1$&$-1$&$\omega_{m}$&$\omega_{m}$&$\omega_{m}$\\ \cline{4-10}
&&&$\leq -2$&$Q_{0}\to +\infty$&$-1$&$-1$&$\omega_{m}$&$\omega_{m}$&$-1$\\
\hline
\parbox[t]{2mm}{\multirow{4}{*}{\rotatebox[origin=c]{90}{Type II}}}& \multirow{2}{*}{$<0$}& \multirow{2}{*}{$<0$}&$\frac{1}{n}$ or $\frac{1}{2}<\frac{a}{b}\neq \frac{1}{n}<1$&$Q_{0}\to +\infty$&$-1$&$-1$&$Q_{1}\to -\infty$&$\omega_{m}$&$\omega_{m}$ \\[2pt] \cline{4-10}
&&&$\frac{1}{m}$ or $0<\frac{a}{b}\neq \frac{1}{m}<\frac{1}{2}$&$Q_{0}\to \mp \infty$&$-1$&$-1$&$Q_{1}\to \pm \infty$&$-1$&$\omega_{m}$\\ [2pt]\cline{2-10}
&\multirow{2}{*}{$>0$}& \multirow{2}{*}{$>0$}&$\frac{1}{n}$ or $\frac{1}{2}<\frac{a}{b}\neq \frac{1}{n}<1$&$Q_{0}\to -\infty$&$-1$&$-1$&$Q_{1}\to +\infty$&$-1$&$-1$\\ [2pt] \cline{4-10}
&&&$\frac{1}{m}$ or $0<\frac{a}{b}\neq \frac{1}{m}<\frac{1}{2}$&$Q_{0}\to \mp \infty$&$-1$&$-1$&$Q_{1}\to \pm \infty$&$-1$&$\omega_{m}$\\ [2pt]
\hline
\parbox[t]{2mm}{\multirow{4}{*}{\rotatebox[origin=c]{90}{Type III}}}& \multirow{2}{*}{$<0$}& \multirow{2}{*}{$<0$}&$\frac{-1}{n}$ or $-1<\frac{-a}{b}\neq \frac{-1}{n}<\frac{-1}{2}$&$-1$&$Q_{0}\to -\infty$&$Q_{0}\to -\infty$&$\omega_{m}$&$Q_{0}\to -\infty$&$Q_{0}\to -\infty$ \\[2pt] \cline{4-10}
&&&$\frac{-1}{m}$ or $\frac{-1}{2}<\frac{-a}{b}\neq \frac{-1}{m}<0$&$-1$&$Q_{0}\to +\infty$&$Q_{0}\to -\infty$&$-1$&$Q_{0}\to +\infty$&$Q_{0}\to -\infty$\\ [2pt]\cline{2-10}
&\multirow{2}{*}{$>0$}& \multirow{2}{*}{$>0$}&$\frac{-1}{n}$ or $-1<\frac{-a}{b}\neq \frac{-1}{n}<\frac{-1}{2}$&$-1$&$Q_{0}\to +\infty$&$Q_{0}\to +\infty$&$-1$&$Q_{0}\to +\infty$&$Q_{0}\to +\infty$\\ [2pt] \cline{4-10}
&&&$\frac{-1}{m}$ or $\frac{-1}{2}<\frac{-a}{b}\neq \frac{-1}{m}<0$&$-1$&$Q_{0}\to +\infty$&$Q_{0}\to -\infty$&$-1$&$Q_{0}\to +\infty$&$Q_{0}\to -\infty$\\ [2pt]
\hline
\parbox[t]{2mm}{\multirow{3}{*}{\rotatebox[origin=c]{90}{Type IV}}}&$>0$&$<0$&$>1$&$Q_{0}\to -\infty$&$-1$&$-1$&$\omega_{m}$&$-1$&$-1$\\ \cline{2-10}
&\multirow{2}{*}{$<0$}& \multirow{2}{*}{$>0$}&$\in (1,2)$&$Q_{0}\to +\infty$&$-1$&$-1$&$\omega_{m}$&$\omega_{m}$&$\omega_{m}$\\ \cline{4-10}
&&&$\geq 2$&$Q_{0}\to +\infty$&$-1$&$-1$&$\omega_{m}$&$-1$&$\omega_{m}$\\
\hline\hline
\end{tabular}\end{center}\label{Table:EoS}
{Note: The quantities $Q_{0}$ and $Q_{1}$ above are as given in Eqs. (\ref{Q0}) and (\ref{Q1}), respectively.}
\end{table}
\subsection*{Acknowledgments}
The authors would like to thank Prof. A. Awad for several discussions. This work is partially supported by the Egyptian Ministry of Scientific Research under project No. 24-2-12.
%
\end{document}